\documentclass[12pt,draftclsnofoot, onecolumn]{IEEEtran}

\normalsize

\usepackage[T1]{fontenc}
\usepackage{color}
\usepackage[protrusion=true,expansion=true]{microtype}

\usepackage{amsmath,amsfonts,amsthm} 
\usepackage[pdftex]{graphicx}	
\usepackage{url}
\usepackage{float}

\usepackage{textcomp}

\usepackage{fancyhdr}
\usepackage{amsthm}
\usepackage{amsmath}
\usepackage{amssymb}
\usepackage{amsmath}
\usepackage{amssymb}
\usepackage{epsfig}
\usepackage{epsf}
\usepackage{amsmath,amsfonts,amsthm,bm} 
\usepackage{subfigure}
\usepackage{graphicx}
\usepackage{url}
\usepackage{color}
\usepackage{enumerate}
\usepackage{algorithm}   
\usepackage[noend]{algpseudocode}
\usepackage{verbatim}
\usepackage{environ}
\usepackage{mathtools}
\DeclarePairedDelimiter\ceil{\lceil}{\rceil}

\usepackage[numbers,sort&compress]{natbib}

\usepackage{enumitem,kantlipsum}

\usepackage{multicol}
\usepackage{tikz}
\usepackage{pgfplots}
\pgfplotsset{compat=1.7}
\usepgfplotslibrary{groupplots} 
\usetikzlibrary{pgfplots.groupplots}

\usepgfplotslibrary{fillbetween}

\usepackage{blindtext}

\usepackage{subfiles} 

\newtheorem*{rep@theorem}{\rep@title}
\newcommand{\newreptheorem}[2]{%
\newenvironment{rep#1}[1]{%
 \def\rep@title{#2 \ref{##1}}%
 \begin{rep@theorem}}%
 {\end{rep@theorem}}}

\newtheorem{theorem}{Theorem}
\newreptheorem{theorem}{Theorem}

\newreptheorem{lemma}{Lemma}

\newtheorem{definition}{Definition}

\newtheorem{remark}{Remark}
\usepackage{url}

\newcommand{\iscomment}[1]{{\color{blue} \bf{{{{IS --- #1}}}}}}    
\newcommand{\sncomment}[1]{{\color{brown} \bf{{{{SN --- #1}}}}}}    

\include{defs}
\begin{document}
\title{Reassembly Codes for\\ the Chop-and-Shuffle Channel
}

%


\author{Sajjad~Nassirpour,~ Ilan~Shomorony,~and~ Alireza~Vahid

\thanks{S. Nassirpour and A. Vahid are with the Department of Electrical Engineering at the University of Colorado, Denver, USA. Email: {\sffamily sajjad.nassirpour@ucdenver.edu} and {\sffamily alireza.vahid@ucdenver.edu}. I. Shomorony is with the Department of Electrical and Computer Engineering at the University of Illinois at Urbana-Champaign, Urbana, USA. Email: {\sffamily ilans@illinois.edu}.}}
\maketitle

\begin{abstract}
We study the problem of retrieving data from a channel that breaks the input sequence into a set of unordered fragments of random lengths, which we refer to as the chop-and-shuffle channel. The length of each fragment follows a geometric distribution. We propose nested Varshamov-Tenengolts (VT) codes to recover the data. We evaluate the error rate and the complexity of our scheme numerically. Our results show that the decoding error decreases as the input length increases, and our method has a significantly lower complexity than the baseline brute-force approach. We also propose a new construction for VT codes, quantify the maximum number of the required parity bits, and show that our approach requires fewer parity bits compared to known results.
\end{abstract}

\begin{IEEEkeywords}
Varshamov-Tenengolts Codes, layered codes, DNA-based data storage, fragmented data, shuffling channel.
\end{IEEEkeywords}

\section{Introduction}
\label{Section:intro}
Recently DNA-based storage has emerged as a promising technology for durable data storage with enormous information density. Several works \cite{church2012next,goldman2013towards,grass2015robust,yazdi2015rewritable,erlich2017dna,organick2018random} have presented DNA-based storage prototypes, which are capable of storing up to 200 Megabytes of data.

DNA contains four nucleotides (Adenine, Cytosine, Guanine, and Thymine), and data can be mapped from the binary domain into the nucleotide domain for storage purposes. However, the DNA strings are exposed to breaks during writing (synthesizing) and reading (sequencing) procedures. During the synthesis process, data is
written onto DNA molecules and stored in a solution where molecules
in the solution are subject to random breaks. As for reading/sequencing, there are two standard methods: (1) {\it Shotgun sequencing} that provides a cost-efficient, high-throughput platform and samples a long DNA molecule with a large number of
short reads. In particular, this method reads the data from random locations of the DNA sequences in a DNA pool
in a parallel fashion~\cite{motahari2013information,yazdi2015rewritable,bornholt2016dna}. However, shotgun sequencing does not create a long sequence, which has potentials for higher density storage and faster reading process; (2) {\it Nanopore sequencing} that enables reading of several thousand bases at a time by moving the DNA string through a small pore with radius in the order of a nanometer~\cite{branton2010potential,bleidorn2016third,jain2016oxford}. However, Nanopore sequencing creates variable-length sequences and increases the error
rate and cost compared to shotgun sequencing. We note that the two common sequencing methods create unordered DNA fragments with fixed or variable lengths, which require reassembly and result in the shuffling channel~\cite{shomorony2020communicating};

Motivated by the above results, in this paper, we consider a chop-and-shuffle channel where the input is a length-$n$ binary data bits (which would be easy to extend to nucleotide domain) that is chopped into a random number, $M$, fragments with lengths $L_1, L_2, \ldots, L_M$ where $L_j, 1\leq j \leq M$ is a random variable with Geometric$(p_n)$ distribution. Here, the channel output is an unordered set of variable-length elements that are shuffled.

The capacity of a chop-and-shuffle channel 
with variable-length fragments is characterized in~\cite{shomorony2020communicating,TornPaperCoding} as
\begin{align}
    \label{eq:capacity}
    C=e^{-\alpha},
\end{align}
where $\alpha = \lim_{n \rightarrow \infty}{p_n \log_2(n)}$, and \cite{shomorony2020communicating,TornPaperCoding} achieve the capacity via random coding. However, random coding is not a practical solution, and in~\cite{TornPaperCoding} a constructive code is also presented but the achievable rates fall short of the capacity.

The main contributions in this paper are as follows:

\begin{itemize}
    \item We devise layered Varshamov-Tenengolts (VT) codes, called ``Nested VT codes,'' to recover the input data from the broken fragments of a chop-and-shuffle channel; 
    \item We focus on scenarios where the channel capacity $C \leq 1$ (i.e., $0 \leq \alpha \leq 1$) and determine the achievable rates of the nested VT codes numerically;  
    \item We study the complexity of our reassembly algorithm in terms of the number of required iterations to recover the input data and illustrate that our method significantly reduces the computational complexity compared to the brute-force approach (which checks all possible permutations);
    \item Finally, we also present a new construction for individual VT codewords with linear complexity that needs fewer parity bits compared to prior results.
\end{itemize}

\vspace{-3mm}

\subsection{Related work}
\label{subsec:related_work}

Rapid advances in DNA-based storage with today's technology attract a significant amount of research attention from an information-theoretic perspective where DNA-based storage acts as a communication channel~\cite{kiah2016codes,yazdi2015rewritable,erlich2017dna,ravi2021coded,shomorony2019capacity,shomorony2021dna,lenz2019coding}. Among these, \cite{shomorony2021dna,lenz2019coding} use coding over a set of short sequences that are randomly sequenced and then shuffled in an unordered fashion. In particular, \cite{shomorony2021dna} studies a storage channel assuming each random sequence is read precisely once, and substitution noise is the only source of error. Then, \cite{lenz2019coding} proposes some error-correcting codes under different scenarios suitable for insertions, deletions, and substitution errors. We note that the above works examine cases with sequences of equal length, while in this paper, we focus on variable-length fragments. 

Moreover, extracting data from a set of its subsequences has been studied in several works \cite{motahari2013information,bresler2013optimal} 
where the data is retrieved from real genome sequences. More precisely, \cite{motahari2013information} shows the minimum number of required reads for
reliable reconstruction and  \cite{bresler2013optimal} demonstrates an optimal assembly algorithm for complete reconstruction. Despite these works assume that the observed sequences overlap, in this paper, we propose an encoding/decoding approach that fits non-overlapping sequences. 

As we described above, the capacity of a chop-and-shuffle channel with variable-length fragments is determined and achieved using random coding in~\cite{shomorony2020communicating}.
Then, \cite{TornPaperCoding} proposes an interleaved-pilot scheme based on de Bruijn sequences to encode and decode the data bits. In particular, \cite{TornPaperCoding} provides a lower-bound on the achievable rate by considering a parity string $\textbf{p}$ with length $n/\zeta$ where $\zeta\geq 2$ is a positive integer and $n$ is the length of encoded bits. Then, it shows that if the length of $j^{\mathrm{th}}$ fragment, $L_j, 1\leq j\leq M$, contains at least $L_j/\zeta$ parity bits in it, the proposed approach allows the decoder to recover the data bits uniquely for long enough $L_j$. However, there is a big gap between the achievable rate of the approach in \cite{TornPaperCoding} and the capacity. As a result, in this paper, we focus on an efficient code based on VT codes that provides higher achievable rate, compared to \cite{TornPaperCoding}. 

Furthermore, in~\cite{nassirpour2020embedded}, we also use the chop-and-shuffle channel to model a DNA-based storage channel where DNA breaking is the main source of error. Then, we present a practical encoding/decoding mechanism based on an embedded structure of VT codes to recover data from DNA fragments (outputs of the chop-and-shuffle channel). However, the results of~\cite{nassirpour2020embedded} are limited to $\alpha =0$ (i.e., capacity is $1$) and do not generalize to $0<\alpha<1$. 
In this manuscript, we present a new solution, Nested VT codes, for $0<\alpha<1$. 


\vspace{-2.5mm}

\subsection{Notations}
\label{subsec:notation}

Table~\ref{table:param} summarizes the notations we use in this paper.

\begin{table}[ht]
\caption{Parameters used in this paper.}
\centering
\begin{tabular}{|c|l||c|l|}
\hline

Parameter & \hspace{0.45in}

Description & Parameter & \hspace{1.1in} Description\\
\hline \hline
$\textbf{d}$ & Input data bits. & $r^{(l)}_i$ & Residue of $\textbf{x}^{(l)}_i$. \\
$n_d$ & Length of $\textbf{d}$.& $L_j$ & Length of $\mathbf{y}_j$. \\
$S(\textbf{x})$ & Set of broken fragments. & $\hat{\textbf{x}}$ & Estimated version of $\textbf{x}$. \\
$d_{sec}$  & Length of one section of $\textbf{d}$. & $\hat{\textbf{x}}^{(l)}_i$ & Estimated version of $\textbf{x}^{(l)}_i$.\\
$\textbf{x}$ & Nested VT codeword. & $\hat{\textbf{d}}^{(l)}_i$ & Estimated version of $\textbf{d}^{(l)}_i$. \\
$\ell$ & Number of layers in $\textbf{x}$. & $\hat{\textbf{d}}$ & A guess of data bits ${\textbf{d}}$.\\
$n$ & Length of $\textbf{x}$. & $M$ & Number of broken fragments.\\
$\textbf{d}^{(l)}_i$ & $i^{\mathrm{th}}$ section of data bits in $l^{\mathrm{th}}$ layer. & m & Number of VT codewords in each layer that create a new\\ 
$\textbf{x}^{(l)}_i$ & VT codeword corresponding to $\textbf{d}^{(l)}_i$.& & data section in the next layer.\\[1ex]  
\hline
\end{tabular}
\label{table:param}

\end{table} 
\vspace{-3.5mm}
\subsection{Outline}

The rest of the paper is organized as follows. In Section~\ref{Section:Problem}, we formulate
the problem. In Section~\ref{Section:Main}, we present the main results of the paper, which include our layered design for VT codes, referred to as nested VT code, a new construction for a single VT code, and the encoding rate of the nested VT code. Then, we provide details on why our encoding/decoding scheme contains an erasure code and the nested VT code in Section~\ref{Section:erasure_nested}. In Sections~\ref{Section:Nested_encoder} and \ref{Section:Nested_decoder}, we explain our proposed encoder and decoder, respectively. We show our results via a numerical example in Section~\ref{Section:simulations}, and section~\ref{Section:conclusion} concludes the paper.
\section{Problem Setting}
\label{Section:Problem}
\begin{figure}
    \centering
    \includegraphics[trim = 0mm 0mm 0mm 0mm, clip, scale=7, width=0.9\linewidth, draft=false]{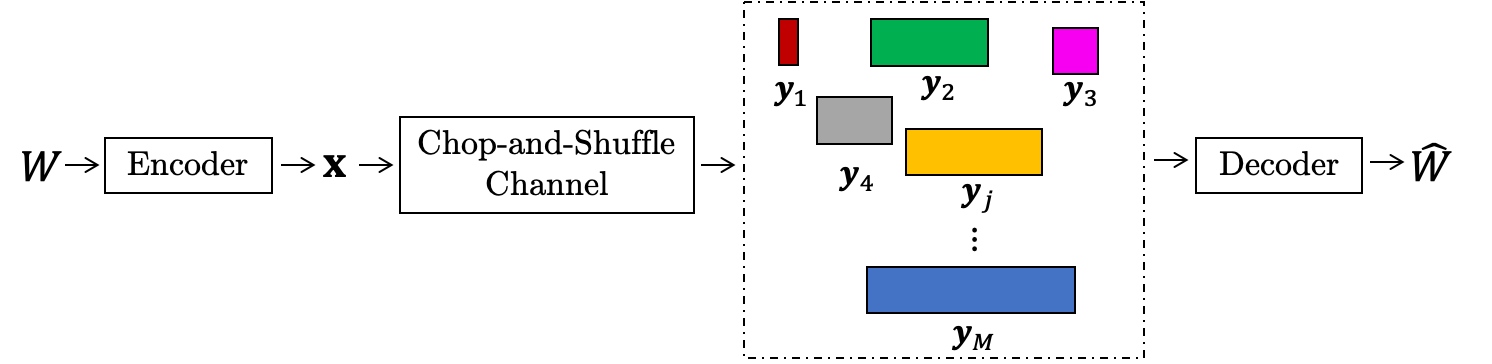}
  \caption{The chop-and-shuffle channel breaks codeword $\textbf{x}$ into $M$ variable-length fragments and then shuffles them where these unordered $M$ fragments are the input of the decoder.\label{Fig:channel_model}}
  \vspace{-6mm}
\end{figure}

Our goal is to devise practical codes for the chop-and-shuffle
channel (CSC) depicted in Fig.~\ref{Fig:channel_model} to reliably communicate an input message $W$ chosen uniformly at random from $\{1,2,...,2^{nR}\}$. In particular, we wish to devise constructive codes to map the input message to a length-$n$ binary codeword $\textbf{x} \in \{0,1\}^n$. 

The chop-and-shuffle channel can be described as follows. The channel first breaks $\textbf{x}$ into $M$ random-length fragments (where $1\le M\le n$ is random itself), and then shuffles these fragments resulting in an unordered set of output fragments. We use $\mathbf{y}_j$, $L_j$, and $S(\textbf{x})=\{\mathbf{y}_1,\mathbf{y}_2,\ldots,\mathbf{y}_M\}$ to denote the $j^{\mathrm{th}}$ output fragment, its corresponding length, and the set of all fragments (i.e., channel output), respectively. We note that $M=1$ implies that the input sequence has gone through the channel perfectly (i.e., no error). 

We assume $L_j$, $1 \leq j \leq (M-1)$, are independently and identically distributed (i.i.d.) random variables having a Geometric distribution as:
\begin{align}
    \label{eq:geometric_ditrb}
    L_j\thicksim \mathrm{Geom}(p_n), \quad 1 \leq j \leq (M-1),
\end{align}
where $0 \leq p_n \leq 1$ captures how frequently codeword $\textbf{x}$ is broken. Since the total length of the fragments should be equal to the length of codeword $\textbf{x}$, we have
\begin{align}
\label{eq:total_length_broken}
    \sum_{j=1}^{M}L_{j}=n,
\end{align}
where the length of the last fragment is determined based on other fragments as
\begin{align}
L_M = n - \sum_{j=1}^{M-1}{L_j}.
\label{length_final_fragment}
\end{align}

Finally, the decoder knows $p_n$ and uses $S(\textbf{x})$ to find an estimate $\hat{W}$ of the input message $W$. An error occurs if $\hat{W}\neq W$, and the average probability of
decoding error is equal to
\begin{align}
    e_n=\mathbb{E}[\Pr\{\hat{W}\neq W\}],
\end{align}
where the expectation is over the random choice of message $W$ and the randomness of the CSC. 
We say that rate $R$ is achievable if there exist encoders and decoders such that $e_n \rightarrow 0$ as $n \rightarrow \infty$. Then, the capacity $C$ is defined as the supremum of all achievable rates.






\section{Main Results}
\label{Section:Main}

In this section, we propose a new layered design for Varshamov-Tenengolts codes that we refer to as ``Nested VT codes,'' which enable recovering data from the output fragments of the chop-and-shuffle channel. In the process, we present a new construction for an individual VT code that suits our designs. Finally, we describe the encoding 
rate of our scheme. Below, we explain the details.

\subsection{Nested VT code}
We first introduce the nested VT code for the chop-and-shuffle channel. VT codes were initially introduced for asymmetric Z-channels in~\cite{varshamov1965codes} as given below.
\begin{definition}
\label{def_VT}
\cite{sloane2002single} For $0\leq r \leq n$, the Varshamov-Tenengolts (VT) code $\textbf{x}$ is a binary encoded string with length $n$ satisfying
\begin{equation}
\label{eq_VT_def}
\sum_{i=1}^{n}ix_i\equiv r \mod (n+1)
\end{equation}
where $x_i$ is the $i^{\mathrm{th}}$ element of $\textbf{x}$ and the sum is evaluated as an ordinary  integer summation, and $r$ is referred to as the residue.

\end{definition}
In~\cite{levenshtein1966binary}, it was shown that VT codes can correct single deletions in the data bits, and in fact, VT codes have been used in deletion channels extensively~\cite{helberg2002multiple,abdel2011helberg,abroshan2019coding,mappouras2019greenflag,vahid2017correcting}. 

In this work, we concatenate and stack individual VT codes with a fixed residue (e.g., $r=0$) 
in several layers to outline our $\ell$-layered nested VT code. 
We use superscript $(l), 1 \leq l \leq \ell$ to denote the parameters in $l^{\mathrm{th}}$ layer of the nested VT code. 

\textbf{Nested VT encoder:} We design our $\ell$-layered encoder to convert the input data bits $\textbf{d}$ with length $n_d$ to codeword $\textbf{x}$ with length $n$. More precisely, our nested VT encoder applies the following steps to create $\textbf{x}$: 
\begin{enumerate}

\item It breaks $\textbf{d}$ into $m^{\ell-1}$ small data sections with length $d_{sec}$ where $m$ represents the number of sections that combine together in each layer. Next, it encodes the $i^{\mathrm{th}}$ data section, $\textbf{d}_{i}^{(1)}, 1\leq i \leq m^{\ell-1}$, to $\textbf{x}_i^{(1)}$, the $i^{\mathrm{th}}$ VT codeword in layer 1 with residue $r_i^{(1)}=0$;

\item Going from layer $l$ to $l+1$, $1 \leq l < \ell$, $m$ VT codewords in layer $l$ are concatenated to form a new data section in for layer $l+1$ ($\mathbf{\textbf{d}}^{(l+1)}_{i}=[\mathbf{\textbf{x}}^{(l)}_i$, $\mathbf{\textbf{x}}^{(l)}_{i+1}$,$\ldots$, $\mathbf{\textbf{x}}^{(l)}_{i+m-1}]$), and then encoded as a new VT code. This reduces the number of sections by a factor of $m$. Finally, the $\ell^{\mathrm{th}}$ layer includes only one VT codeword.





\end{enumerate}

\begin{figure}[!ht]
    \centering
    \includegraphics[trim = 0mm 0mm 0mm 0mm, clip, scale=6, width=0.75\linewidth, draft=false]{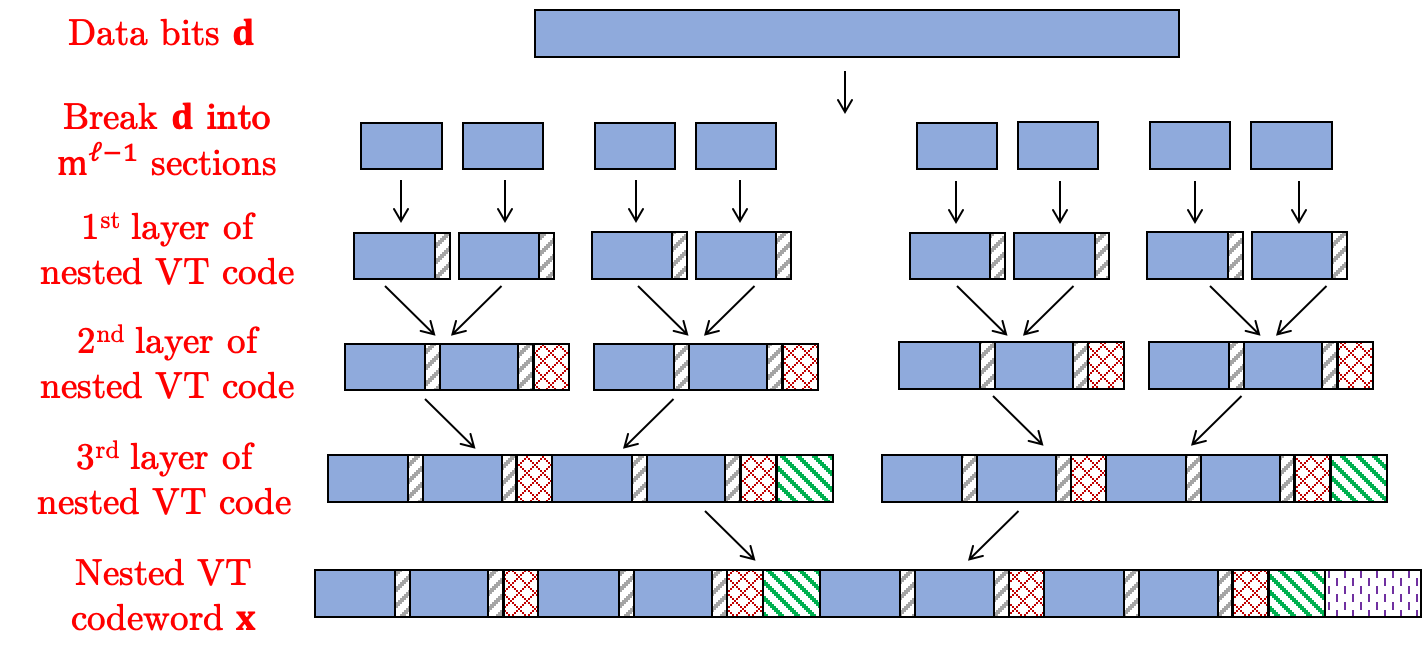}
  \caption{ Nested VT encoder with four layers and $m=2$.\label{Fig:nested_VT_encoder_pattern}}
  \vspace{-6mm}
\end{figure}

Fig.~\ref{Fig:nested_VT_encoder_pattern} depicts an example of the nested VT encoder with $\ell=4$ and $m=2$.  We will explain the technical details of the nested VT encoder in Section \ref{Section:Nested_encoder}.

We note that $\ell$ and $m$ are design parameters that determine the attained rate and the average error. We will derive upper and lower bounds on the rate associated with the nested VT code in Theorem~\ref{theorem:encoding_rate}. 

\textbf{Nested VT decoder:} The output of the chop-and-shuffle channel is $M$ different fragments with variable lengths. Here, our goal is to create a combination of these fragments that satisfies all VT conditions in different layers. We name this combination $\hat{\textbf{x}}$ as an estimated version of $\textbf{x}$, and then remove the parity bits from $\hat{\textbf{x}}$ to recover $\hat{\textbf{d}}$. 

We provide a high-level overview here and defer the technical details to Section~\ref{Section:Nested_decoder}. We will use ``candidate combinations'' and  ``helpful fragments'' to provide this high-level description. We say a combination of fragments is considered as a candidate combination if it satisfies all possible VT conditions in different layers with respect to its length. We let the empty set to be a candidate combination as well. Additionally, a fragment is called helpful if integrating that fragment with a candidate combination results in a new candidate combination.

The decoder begins its task by searching among all broken fragments to find the helpful fragment(s) and saves helpful fragments as candidate combinations. Then, it uses other fragments to find new helpful fragments and combines them with the candidate combination(s) to make a longer and longer candidate combinations. We continue this procedure until we attain $\hat{\textbf{x}}$ of length $n$ 
that meets all VT conditions. Finally, the decoder removes the parity bits to obtain $\hat{\textbf{d}}$. 
We note that our nested VT decoder may find two or more candidate combinations of length $n$ that meet all the VT conditions. We will explain later in Section~\ref{subsect:VTdecoder_erasure_decoder} that we will use the overlap of these candidates plus an erasure code to recover the data.

To contain complexity and simulation runtime, we allow the decoding algorithm to run for a limited time, $\Delta$, an declare an error if the decoder obtains at least one output string during $\Delta$ and $\hat{\textbf{d}} \neq \textbf{d}$. In general, there will be some cases where our algorithm cannot find $\hat{\textbf{d}}$ within the given time. In Section~\ref{Section:simulations}, we show that these cases represent a small fraction of the total cases and they decrease as we increase $\Delta$. 

A natural question is whether it would be beneficial to consider other possibilities for $r_i^{(l)}$ such as using $r_i^{(l)}\neq 0$ or assigning different residues to each VT codeword within the nested structure. We will study this in Section~\ref{Section:simulations} and show that: (i) with $r_i^{(l)}\neq 0$, the decoding error increases; (ii) using different residues provides almost the same error rate, but slightly decreases the number cases where our algorithm cannot find $\hat{\textbf{d}}$ within $\Delta$.

\subsection{A new construction for a single VT Code}
\label{subsect:generate_VT_code}

In \cite{sloane2002single}, the author describes a VT encoder that generates codeword $\textbf{x}$ from data bits $\textbf{d}$ with linear complexity. More precisely, \cite{sloane2002single} constructs VT codeword $\textbf{x}$ as 
\begin{align}
\label{eq:construct_x}
\textbf{x}=[\textbf{d}~~\textbf{p}],
\end{align}
where $\textbf{d}$ is the data bits with length $n_d$, $\textbf{p}$ represents the parity bits with length $p$, and $p$ equals to
\begin{align}
\label{eq:ParityLinear}
    p=\Bigg\lceil\sqrt{2n_d+\frac{9}{4}}+\frac{1}{2}\Bigg\rceil,
\end{align}
where $\ceil{.}$ is the ceiling function.

We note that~\cite{sloane2002single} does not provide a proof for this claim. 


In this paper, we propose a new encoding method for VT codes with a lower number of required parity bits compared to \eqref{eq:ParityLinear}. The following theorem quantifies the number of parity bits in our method. 

\begin{theorem}
\label{theorem_length_parity}
Given any data-bit sequence of length $n_d$, it is possible to construct a VT code as defined in (\ref{eq_VT_def}) of length $n$ with residue $0 \leq r \leq n$ where $n = n_d + p$ for $p=\Big\lceil\frac{1+\sqrt{1+8n_d}}{2}\Big\rceil$.

\end{theorem}

\begin{proof}
The details are described in Appendix~\ref{appndx:thm_length_parity}.
\end{proof}

We provide a detailed description of how our VT encoder obtains parity bits $\textbf{p}$ with linear complexity in Appendices~\ref{appndix:find_parity}  and \ref{appndix:complexity_parity}. 


\subsection{Encoding rate of the nested VT code}
\label{subsect:encoding_rate}
As we described above, in our proposed nested VT encoder, the lengths of input data bits $\textbf{d}$ and codeword $\textbf{x}$ are equal to $m^{(\ell-1)}d_{sec}$ ($\textbf{d}$ includes $m^{(\ell-1)}$ data sections of length $d_{sec}$) and $n$, respectively. Hence, the encoding rate of our nested VT code is
\begin{align}
\label{eq_encoding_rate_def}
    R=\frac{m^{(\ell-1)}d_{sec}}{n}\overset{(a)}{=}\frac{m^{(\ell-1)}d_{sec}}{\Tilde{n}_{\ell}},
\end{align}

where $\Tilde{n}_{\ell}$ represents the length of a VT codeword in the $\ell^{\mathrm{th}}$ layer, and $(a)$ holds true because the last layer of the nested VT code includes one codeword.


Moreover, we know that each data section in layer $l$, $2 \leq l \leq \ell$, contains $m$ different VT codewords of the $(l-1)^{\mathrm{th}}$ layer. Thus, we use Theorem~\ref{theorem_length_parity} and the length of each data section in layer $l$ to determine the length of each VT codeword in layer $l$. As a result, we need the length of codewords in all layers to attain $\Tilde{n}_{\ell}$, which gives us a recursive formula. Below, we present the details.

The first layer of our encoder includes $m^{(\ell-1)}$ data sections where the length of each section is equal to $d_{sec}$. Therefore, from Theorem~\ref{theorem_length_parity}, we have
\begin{align}
\label{tilda_n_1_main}
    \Tilde{n}_1=d_{sec}+\Bigg\lceil\frac{1+\sqrt{1+8d_{sec}}}{2}\Bigg\rceil,
\end{align}
where $\Tilde{n}_1$ is the length of each codeword in layer 1.

In the second layer, each data section consists of $m$ different VT codewords of the first layer. Hence, we calculate $\Tilde{n}_2$ (the length of each codeword in layer 2), as
\begin{align}
\label{eq:n2_tilde_main}
   \Tilde{n}_2=m\Tilde{n}_1+\Bigg\lceil\frac{1+\sqrt{1+8m\Tilde{n}_1}}{2}\Bigg\rceil. 
\end{align}
Following the same rule, we have
\begin{align}
\label{eq:nl_tilde_main}
   \Tilde{n}_{\ell}=m\Tilde{n}_{\ell-1}+\Bigg\lceil\frac{1+\sqrt{1+8m\Tilde{n}_{\ell-1}}}{2}\Bigg\rceil. 
\end{align}

The following theorem
provides lower and upper bounds of the encoding rate of our nested VT code.

\begin{figure}
    \centering
    \includegraphics[trim = 0mm 0mm 0mm 0mm, clip, scale=5, width=0.99\linewidth, draft=false]{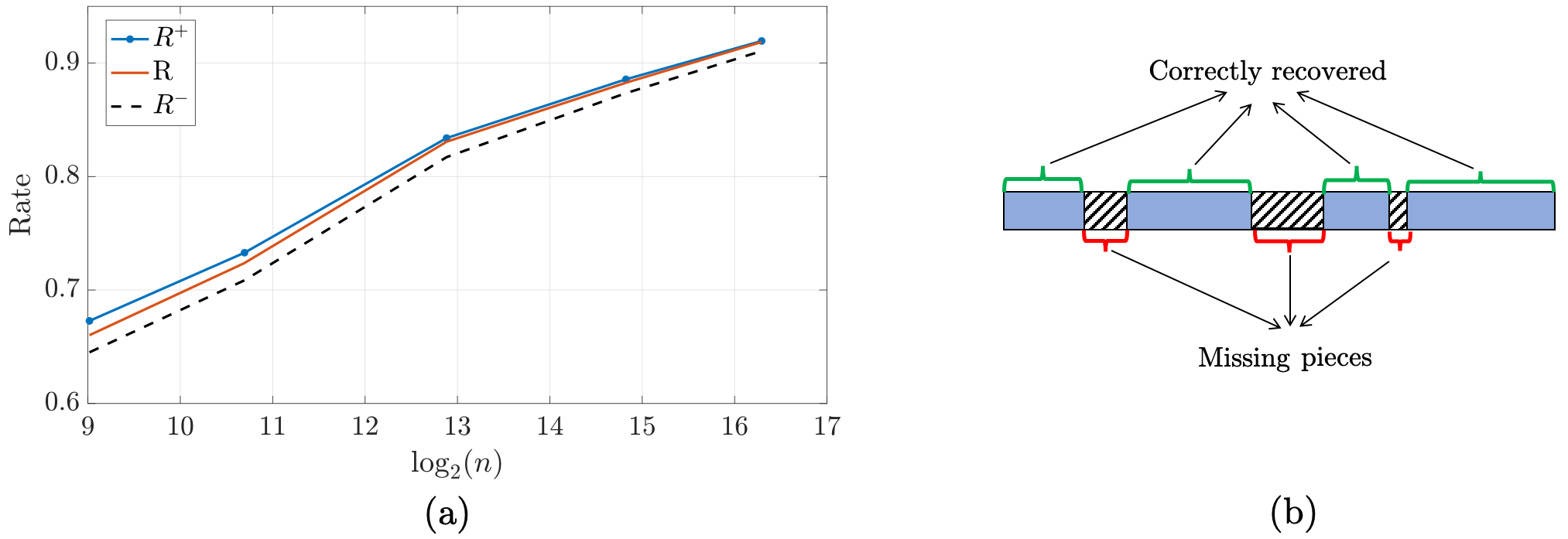}
  \caption{(a) The lower and upper bounds of the nested VT code encoding rate versus the actual encoding rate of nested VT code; (b) If there are some fragments of very short length, the nested VT code fails to recover all data bits.\label{Fig:encoding_rate_boundaries}
  }
  \vspace{-8mm}
\end{figure}

\begin{theorem}
\label{theorem:encoding_rate}
Given that $n_d=m^{(\ell-1)}d_{sec}$ is the length of data bits $\textbf{d}$ and for $d_{sec}\geq 36$, the encoding rate of the $\ell$-layered nested VT code is bounded by 
\begin{align}
\label{eq:nested_rate_main_results}
R^-<R<R^+,
\end{align}
where 
\begin{align}
\label{eq:R_minus}
    R^-=\frac{m^{\ell-1}d_{sec}}{\Big(m^{(\frac{\ell-1}{2})}\sqrt{d_{sec}}+\frac{\sqrt{2.5}}{2}\frac{m^{\frac{\ell}{2}}-1}{\sqrt{m}-1}\Big)^2},
\end{align}
and the upper bound, $R^+$, is equal to
\begin{align}
\label{eq:R_plus}
    R^+=\frac{2m^{\ell-1}d_{sec}}{\Big(m^{(\frac{\ell-1}{2})}\sqrt{2d_{sec}}+\frac{m^{\frac{\ell}{2}}-1}{\sqrt{m}-1}\Big)^2}.
\end{align}
\end{theorem}
\begin{proof}
We defer the proof to Appendix~\ref{appndix_proof_encoding_rate}.
\end{proof}
In Fig.~\ref{Fig:encoding_rate_boundaries}(a), we describe how $R^-$, $R$, and $R^+$ behave when $\ell=3$ and $9\leq \log_2(n)<17$. Here, the encoding rate approaches $R^+$ as $n \to \infty$. 
In general, we select $\ell$ based on $\alpha$, and then we let $d_{sec}\to\infty$ and $m\to\infty$ to capture $n\to\infty$. Specifically, in Fig.~\ref{Fig:encoding_rate_boundaries}(a), we set $\ell=3$ and increase $d_{sec}$ and $m$ to get the results. 
It is worth mentioning that $R$ describes the encoding rate and not the achievable rate. Of course, to have any hope of having vanishing error probability, we need $R < C$. Therefore, we always select $\ell$, $m$, and $d_{sec}$ such that $R < e^{-\alpha}$.





\section{Erasure code and nested VT code}
\label{Section:erasure_nested}
Ideally, we would like to recover $\textbf{x}$ perfectly; however, for various reasons we may not be able to obtain a unique sequence as the output of the decoder. 
A simple reason would be some of the fragments could fit at different positions, and this results in multiple reassembled sequences as the outputs of the nested VT decoder that satisfy all VT conditions. While this may seem like an error at first glance, we take the overlap of all the outputs as the final reassembled sequence, and our simulations show this, if treated with care, will result in a correct reconstruction with high probability. Of course, taking the overlap results in some missing pieces as Fig.~\ref{Fig:encoding_rate_boundaries}(b), which can be viewed as erasures.



As a result, to handle these erasures, we first apply an erasure code to the data bits 
before feeding the sequence to the nested VT encoder. We note that since our nested VT code preserves the data structure (i.e., it places parity bits at the end of each codeword), we can apply the erasure code before our proposed encoder and data recovery will require an erasure decoder at the end. This way, the total rate of our scheme with erasure and nested VT codes is equal to 
\begin{align}
    R_t=(1-\epsilon)R,
\end{align}
where $(1-\epsilon)$ is the erasure code rate, and we will determine $\epsilon$ heuristically with simulations.



\begin{figure}
    \centering
    \includegraphics[trim = 0mm 0mm 0mm 0mm, clip, scale=5, width=0.65\linewidth, draft=false]{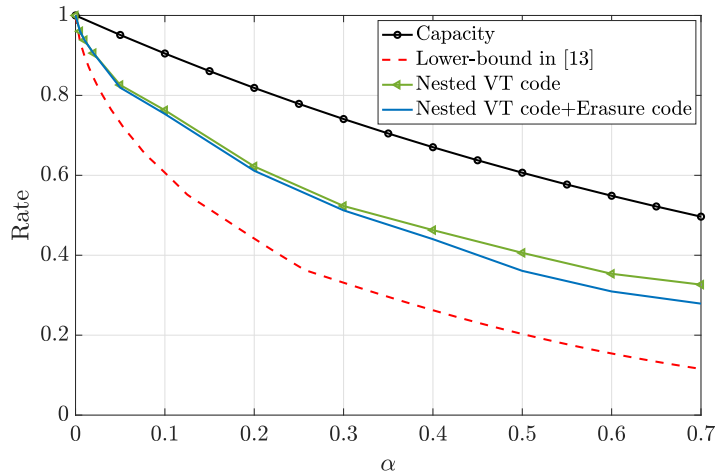}
  \caption{Comparison between the rate of nested VT code+erasure code,  lower-bound of \cite{TornPaperCoding}, nested VT code, and the capacity.\label{Fig:rate_our_vs_nested}}
\end{figure}

As we mentioned in Section~\ref{Section:Main}, we need $R<C$ as a necessary condition to expect the error probability to decrease as $n$ increases in the chop-and-shuffle channel; thus, in our encoding scheme, we select $\ell$, $m$, and $d_{sec}$ such that $R < e^{-\alpha}$. Fig.~\ref{Fig:rate_our_vs_nested} compares the rate of the nested VT code plus erasure code with the baseline nested VT code where the value of $\epsilon$ is obtained heuristically. Furthermore, in Fig.~\ref{Fig:rate_our_vs_nested}, we compare the rate of our approach (i.e., erasure code + nested VT code) with the capacity and the lower-bound derived in~\cite{TornPaperCoding} using de Bruijn sequences. We observe that our approach provides a higher rate compared to the interleaved-pilot scheme of \cite{TornPaperCoding}.


\section{Proposed encoder}
\label{Section:Nested_encoder}
\begin{figure}
    \centering
    \includegraphics[trim = 0mm 0mm 0mm 0mm, clip, scale=5, width=0.6\linewidth, draft=false]{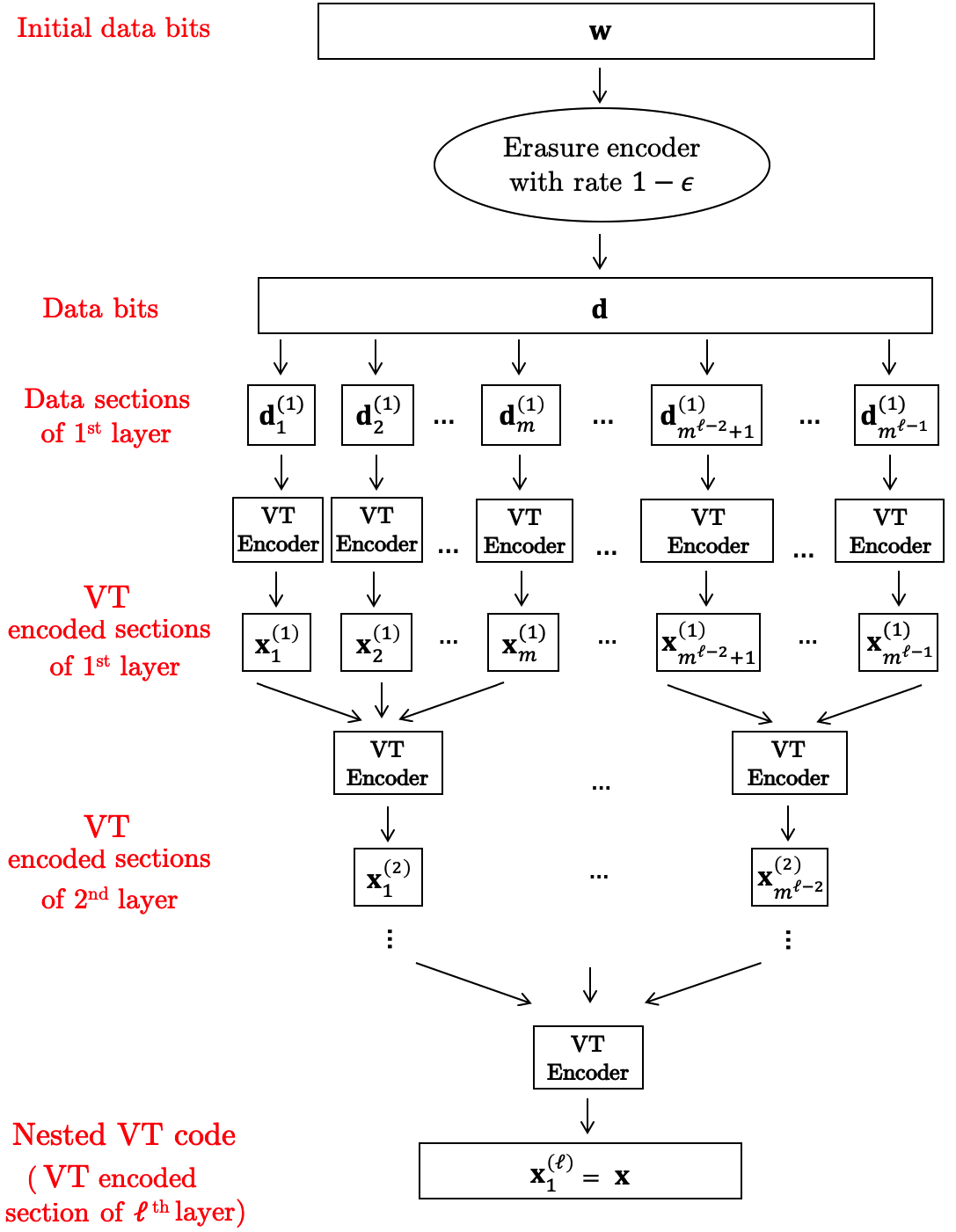}
  \caption{Our encoding structure includes both erasure and nested VT codes.
  \label{Fig:nested_encoder_scheme}}
  \vspace{-7mm}
\end{figure}

In this part, we present our proposed nested VT encoder with $\ell$ layers. 
As we discussed in Section~\ref{Section:erasure_nested}, we need an erasure code before our nested VT code to compensate for the missing pieces at the end. Therefore, we build our encoder by concatenating an erasure encoder to the nested VT encoder as in Fig.~\ref{Fig:nested_encoder_scheme}.

\noindent \textbf{Initial processing}: Suppose $\textbf{w}$ is the binary input sequence. We apply an erasure code with the rate $0 \leq (1-\epsilon) \leq 1$ to $\textbf{w}$ to obtain $\textbf{d}$ as the input sequence to our nested VT encoder. For simplicity, we will refer to $\textbf{d}$ as data bits. As mentioned before, the rate for the erasure code will be determined heuristically;



\noindent $\textbf{Breaking the input data bits d into~} \mathbf{m^{\ell-1}~}\textbf{sections:}$ The nested VT encoder breaks $\textbf{d}$ with length $n_d=m^{(\ell-1)}d_{sec}$ into $m^{(\ell-1)}$ data sections where the length of each data bits section is equal to $d_{sec}$; 

\noindent $\textbf{Encode each data section:}$ In this step, we apply VT code on each data section $\textbf{d}^{(l)}_i, 1 \leq l \leq \ell, 1 \leq i \leq m^{\ell-l}$ to obtain VT codeword $\textbf{x}^{(l)}_i$. To do this, in the $i^{\mathrm{th}}$ section of data bits in layer $l$, we use residue $r_i^{(l)}=0, 1 \leq i \leq m^{(\ell-l)}$ to encode data section $\textbf{d}^{(l)}_i$. 

\noindent $\textbf{Combining the codewords:}$ In layer $l, 2 \leq l \leq \ell$, we merge every $m$ VT codewords of layer $(l-1)$ to create a new data section. Specifically, we set 
\begin{align}
    \textbf{d}_i^{(l)}=[\textbf{x}_{i}^{(l-1)}, \textbf{x}_{i+1}^{(l-1)}, \ldots, \textbf{x}_{i+m-1}^{(l-1)}], \quad 1 \leq i \leq m^{\ell-l}.
\end{align}
We use these new sections as the input of the VT encoder in layer $l$. This way, the number of data bits sections is reduced by a factor of $m$ in each layer. Then, we use the same approach to encode the data-bit sections in the new layer. The VT codeword of layer $\ell$ will be the output of the nested VT encoder. 
   
Moreover, we present a pseudo code in Algorithm~\ref{algo:algo_encode} to describe how our proposed encoder runs the encoding procedure.
\begin{algorithm}
  \caption{Encoding Algorithm}
    \label{algo:algo_encode}
  \begin{algorithmic}[1]
  \State Attain $\textbf{d}$ applying an erasure code with rate $1-\epsilon$ to $\textbf{w}$;\
  \State Break $\textbf{d}$ into $m^{(\ell-1)}$ sections as $\textbf{d}^{(1)}_1,\textbf{d}^{(1)}_2, \ldots, \textbf{d}^{(1)}_{m^{\ell-1}}$;\
  \For{$l\in [1,\ell]$}
    \For{$i\in [1,m^{(\ell-l)}]$}
        \State Encode $\textbf{d}^{(l)}_i$ to a VT codeword as $\textbf{x}^{(l)}_i$ based on residue $r_i^{(l)}=0$;\
    \EndFor
    \If{$l\neq \ell$}
    \State Merge every $m$ codewords in layer $l$ to create a new data-bit section in layer $l+1$;\
    \EndIf
  \EndFor
  \State Consider $\textbf{x}=\textbf{x}_1^{(\ell)}$ as the nested VT codeword.
\end{algorithmic}
\end{algorithm}
First, we use an erasure encoder and convert initial data bits $\textbf{w}$ into $\textbf{d}$. In the second line of Algorithm~\ref{algo:algo_encode}, we break the input data bits $\textbf{d}$ into $m^{(\ell-1)}$ sections. 
Next, in each encoding layer, we do the following tasks: (1) We encode each $\textbf{d}^{(l)}_i$ to VT codeword $\textbf{x}^{(l)}_i$ based on a fixed residue ($r_i^{(l)}=0$); (2) If $l \neq \ell$, we combine every $m$ VT codewords together to create a new data-bit section for the next layer. Finally, we define the output of the last layer as the nested VT codeword $\textbf{x}$.

Notice that although we design the nested VT code by deploying $r_i^{(l)}=0$ for all codewords, one possible question is what occurs if we use other methods such as $r_i^{(l)}\neq 0$ or different unique residues as $r_i^{(l)}=i-1$ for each codeword in each layer? To answer this, we will compare the decoding error of using $r_i^{(l)}=0$, $r_i^{(l)} \neq 0$, and different residues in Section~\ref{Section:simulations}. We will describe that using different residues slightly reduces the number of cases where the decoder fails to find a solution during limited time $\Delta$, and using the same {\it non-zero} residue would increase the error rate. 
\vspace{-6mm}
\subsection{An example of the nested VT encoder}
\begin{figure}
    \centering
    \includegraphics[trim = 0mm 0mm 0mm 0mm, clip, scale=5, width=0.65\linewidth, draft=false]{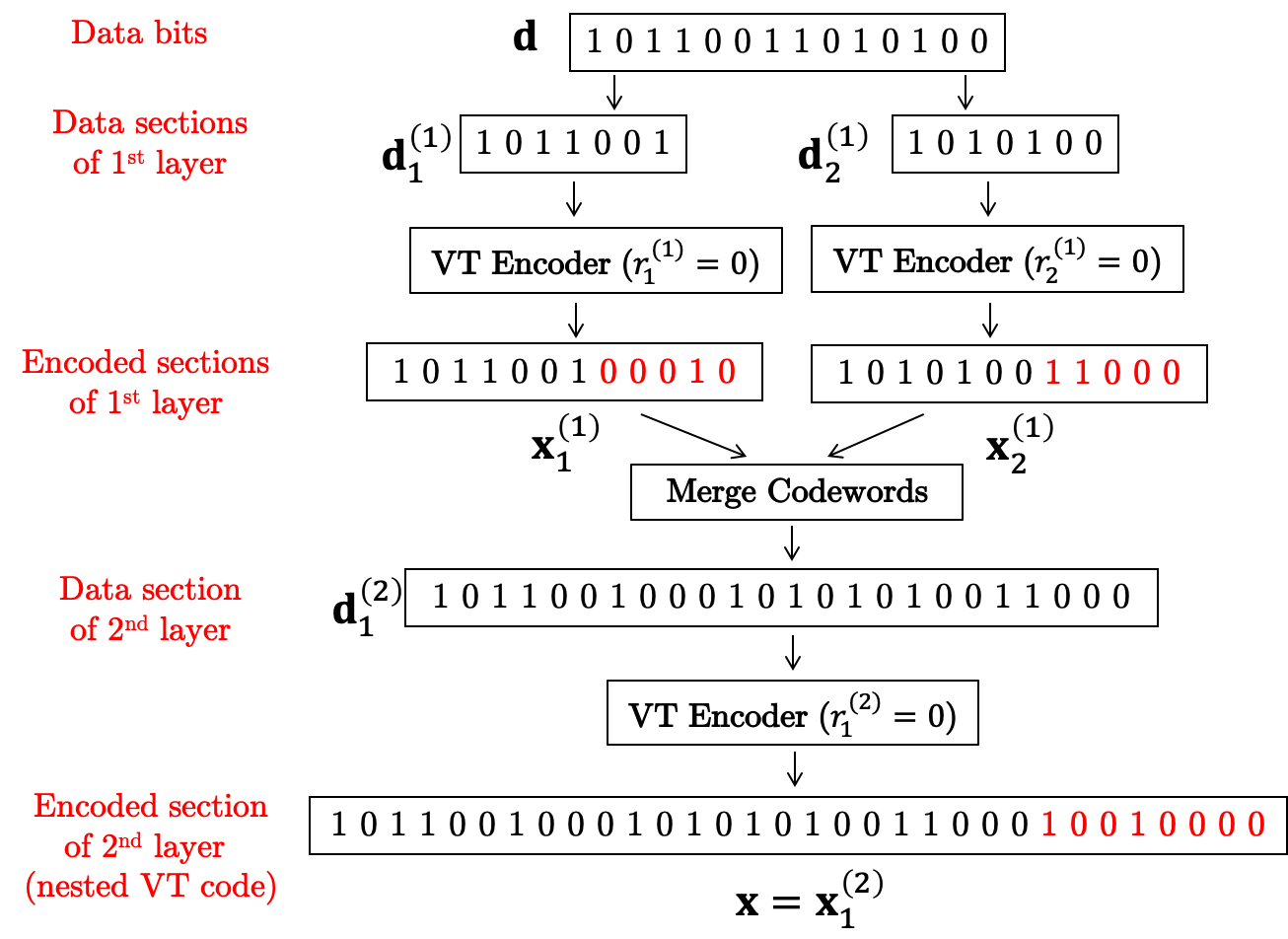}
  \caption{An example of nested VT encoder when $\ell=2$, $m=2$, and $d_{sec}=7$.\label{Fig:ex_nested_encoder}}
   \vspace{-7mm}
\end{figure}

We provide an example in Fig.~\ref{Fig:ex_nested_encoder} with $\ell=2$ layers and $m=2$ to illustrate our nested VT encoder. Here, we focus on the nested VT encoder and remove the erasure encoder to simplify the descriptions. We consider $\textbf{d}=[1~0~1~1~0~0~1~1~0~1~0~1~0~0]$ as the input data bits with $n_d=14$ and $d_{sec}=7$. To encode the data bits, we break $\textbf{d}$ into $m^{\ell-1}=2$ sections as $\textbf{d}_1^{(1)}=[1~0~1~1~0~0~1]$ and $\textbf{d}_2^{(1)}=[1~0~1~0~1~0~0]$. Then, we apply a VT code with $r_1^{(1)}=0$ to $\textbf{d}_1^{(1)}$ to obtain $\textbf{x}_1^{(1)}=[1~0~1~1~0~0~1~0~0~0~1~0]$. Similarly, we get $\textbf{x}_2^{(1)}=[1~0~1~0~1~0~0~1~1~0~0~0]$ using $r_2^{(1)}=0$. In the second layer, we merge $\textbf{x}_1^{(1)}$ and $\textbf{x}_2^{(1)}$ to construct $\textbf{d}_1^{(2)}=[1~0~1~1~0~0~1~0~0~0~1~0~1~0~1~0~1~0~0~1~1~0~0~0]$ as the only data-bit section in that layer. Finally, we apply a  VT code with residue $r_1^{(2)}=0$ to attain $\textbf{x}=\textbf{x}_1^{(2)}=[1~0~1~1~0~0~1~0~0~0~1~0~1~0~1~0~1~0~0~1~1~0~0~0~1~0~0~1~0~0~0~0]$ as the nested VT codeword. 

\section{Proposed decoder}
\label{Section:Nested_decoder}

To explain how our proposed decoder works, we first look at an idealized decoder where the decoding algorithm finds only one permutation of the fragments as the recovered codeword. We also provide an example to demonstrate the details. Then, we discuss how we incorporate erasure codes in non-ideal scenarios. Further details are provided below.
\vspace{-5mm}
\subsection{Ideal nested VT decoder}
\begin{figure}
    \centering
    \includegraphics[trim = 0mm 0mm 0mm 0mm, clip, scale=5, width=0.78\linewidth, draft=false]{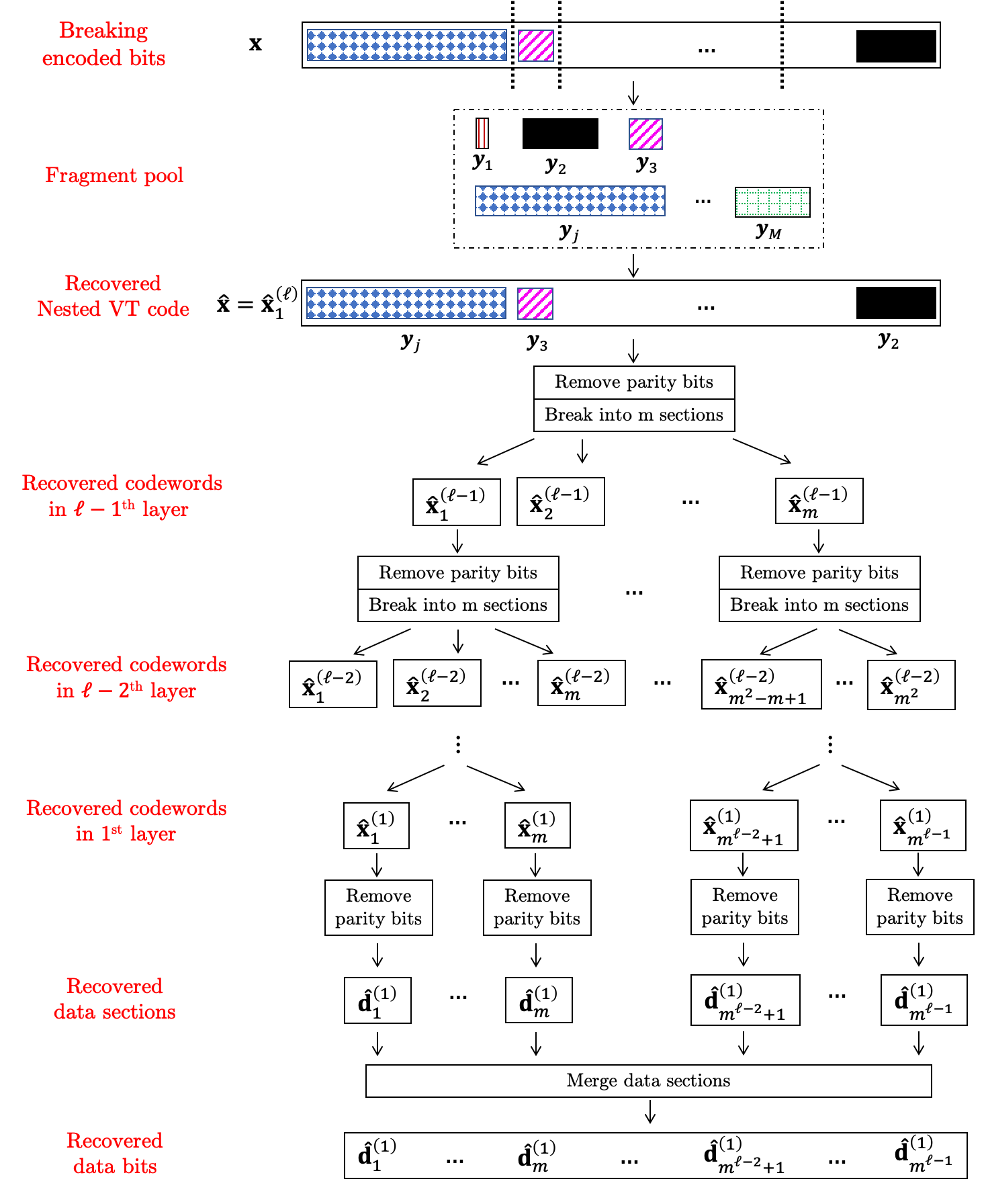}
  \caption{Ideal nested VT decoder\label{Fig:nested_decoder}}
  \vspace{-5mm}
\end{figure}

In this part, we focus on an ideal nested VT decoder as shown in Fig.~\ref{Fig:nested_decoder}, which reconstructs only one permutation of the fragments as the recovered codeword. We assume the chop-and-shuffle channel breaks codeword $\textbf{x}$ into $M$ fragments with random lengths and then shuffles these fragments. To track the fragments during the process, we allocate $M$ labels as $\mathbf{y}_j, j=1,2,\ldots, M$ to the fragments. Before describing in detail, we define some parameters to be used in decoding.

\textbf{Matrix $\mathbf{N}$:} This matrix represents the position of the last bit in all VT codewords in all layers. The size of this matrix is $\ell \times m^{(\ell-1)}$ where $l^{\mathrm{th}}$ row has $m^{(\ell-l)}$ non-zero elements, corresponding to $m^{(\ell-l)}$ codewords in layer $l$, and $N_{l,i}, 1\leq l \leq \ell, 1\leq i \leq m^{(\ell-l)}$ shows the position of the last bit in $\textbf{x}_i^{(l)}$. For example, suppose $d_{sec}=24$, $m=2$, and $\ell=4$, then matrix $\mathbf{N}$ is equal to
\begin{align}
\label{eq:matrix_N}
    \mathbf{N}=
\begin{bmatrix}
32 &  64 & 108 & 140 & 202 & 234 & 278 & 310 \\
76 &  152 & 246 & 322 & 0 & 0 & 0 & 0 \\170 &  340 & 0 & 0 & 0 & 0 & 0 & 0 \\
367 & 0 & 0 & 0 & 0 & 0 & 0 & 0 \\
\end{bmatrix},
\end{align}

where $32$ is the length of $\textbf{x}^{(1)}_1$ that is obtained using Theorem~\ref{theorem_length_parity} and $d_{sec}=24$. Furthermore, we know that $\textbf{x}^{(1)}_2$ is the next codeword in $\textbf{x}$ with length $32$; therefore, the position of the last bit in $\textbf{x}^{(1)}_2$ is equal to $32+32=64$. Then, since $m=2$, every two codewords in layer 1 are merged together to create a new data bits section in layer 2. Thus, $\textbf{d}^{(2)}_1=[\textbf{x}^{(1)}_1~~ \textbf{x}^{(1)}_2]$ with length $64$, and using $\textbf{d}^{(2)}_1$ and Theorem~\ref{theorem_length_parity} leads to have $\textbf{x}^{(2)}_1$ with length $76$. We can get other numbers in matrix $\mathbf{N}$ by following the same rule.


\textbf{Candidate combination:} Suppose $\boldsymbol\Gamma$ is a sequence that can be generated from the elements of $S(\textbf{x})$ and $|\boldsymbol\Gamma|$ is the size of $\boldsymbol\Gamma$ where $0\leq |\boldsymbol\Gamma|\leq M$. Here, we use $\gamma$ to denote the length of $\boldsymbol\Gamma$. More precisely, we have
\begin{align}
    \gamma=\sum_{j^{\prime}=1}^{|\boldsymbol\Gamma|} L_{j^{\prime}}, \quad \mathbf{y}_{j^{\prime}}\in \boldsymbol\Gamma,
\end{align}
where $L_{j^{\prime}}$ is the length of $\mathbf{y}_{j^{\prime}}$. Notice that $\gamma=0$ if $|\boldsymbol\Gamma|=0$. Then, we check the VT condition of every codeword $\textbf{x}^{(l)}_i, 1\leq l \leq \ell, 1\leq i \leq m^{(\ell-l)}$ corresponding to element $N_{l,i}$ in matrix $\mathbf{N}$ if $N_{l,i}\leq \gamma$. Finally, we call $\boldsymbol\Gamma$ a candidate combination if it satisfies VT conditions for all the above codewords. 


We demonstrate an example to make the definition of a candidate combination clear. Consider $d_{sec}=24$, $m=2$, $\ell=4$, and $\boldsymbol\Gamma$ is a sequence of some fragments in $S(\textbf{x})$ with length $\gamma=153$. By using matrix $\mathbf{N}$ in \eqref{eq:matrix_N}, we find that $N_{1,1}=32, N_{1,2}=64, N_{1,3}=108, N_{1,4}=140, N_{2,1}=76$, and $N_{2,2}=152$ are smaller than $\gamma=153$. Therefore, if $\boldsymbol\Gamma$ meets VT conditions of $\textbf{x}^{(1)}_1,\textbf{x}^{(1)}_2, \textbf{x}^{(1)}_3, \textbf{x}^{(1)}_4, \textbf{x}^{(2)}_1$, and $\textbf{x}^{(2)}_2$, it will be a candidate combination.

\begin{remark}
We note that any sequence that is generated from the elements of $S(\textbf{x})$ with a length less than $N_{1,1}$ (position of the last bit in $\textbf{x}_1^{(1)}$) is also a candidate combination because it includes no codeword and needs to wait for other fragments in future steps.
\end{remark}
\begin{remark}
For consistency, we consider $\phi$ (empty sequence) as a candidate combination in the first round of decoding. 
\end{remark} 
\textbf{Helpful fragment} We say $\mathbf{y}_j$ with length $L_j$ is a helpful fragment if the combination of $\mathbf{y}_j$ and any candidate combination creates a new candidate combination. 

\textbf{Limited memory:} As $\alpha$ increases, the number of fragments increases and the length of fragments decreases. Thus, our decoder requires to check a fast-growing number of possible cases, resulting in higher computational complexity. To tackle this issue, we define a limited memory with parameter $\tau \in \mathbb{Z}^+$. After every $\tau$ iterations, our algorithm keeps the candidate combinations with the highest number of satisfied VT conditions. If the algorithm cannot resume with any of the kept candidate combinations in the subsequent iterations, it increases $\tau$ by one and repeats the process.  Our simulation results show that this method reduces the complexity of our approach significantly compared to a brute-force search, see Section~\ref{Section:simulations}. 

As a high-level overview, our decoder runs multiple rounds of a searching algorithm between all fragments to find $\mathbf{\hat{\textbf{x}}}$, which satisfies all VT conditions. In the first round, it searches through all fragments to get helpful fragment(s) and consequently candidate combinations. Then, the decoder seeks to detect new helpful fragment(s) to integrate with candidate combinations to create new longer candidate combinations. Finally, the decoder finds $\mathbf{\hat{\textbf{x}}}$ if it is a candidate combination with length $n$. 

When our decoder obtains $\mathbf{\hat{\textbf{x}}}$, the rest of the decoding procedure will be simple. In particular, the decoder removes the parity bits from $\mathbf{\hat{\textbf{x}}}$ to get $\mathbf{\hat{\textbf{d}}}$. In Algorithm~\ref{algo:decoder}, we demonstrate how our ideal nested VT decoder performs the decoding procedure.



\begin{algorithm}
  \caption{Decoding Algorithm}
  \label{algo:decoder}
  \begin{multicols}{2}
  \begin{algorithmic}[1]
  \State $\tau=1$;\
  \State $K=1$;\
  \State $\boldsymbol\Gamma_1=\phi$;\
  \State $\beta_1=S(\textbf{x})$;\
  \State $\tau^{\prime}=0$;\
  \State $Rep=0$;\
  \For{$k \in [1,K]$}{
  \For{$\mathbf{y}_j \in \beta_k$}{
  \State $\boldsymbol\Gamma^{\prime}=[\boldsymbol\Gamma_k ~~\mathbf{y}_j]$;\
  \If{$\boldsymbol\Gamma^{\prime}$ is a candidate combination}
  \State $Rep \leftarrow Rep+1$;
  \State $\boldsymbol\Gamma_{K+Rep}=\boldsymbol\Gamma^{\prime}$;\
  \State $\beta_{K+Rep}=$ Fragments $\in \{S(\textbf{x})\setminus$ {\color{white}123456789}$\boldsymbol\Gamma_{K+Rep}\}$;\ 
  \If{Length $\boldsymbol\Gamma_{K+Rep}=n$}
  \State $\hat{\textbf{{x}}}=\boldsymbol\Gamma_{K+Rep}$;\
  \State Go to line 29;\
  \EndIf
  \EndIf
  }
  \EndFor
  }
  \EndFor
  \For{$k \in [1,K]$}
  \State Remove $\boldsymbol\Gamma_k$;\
  \EndFor
  \State $\tau^{\prime}\leftarrow\tau^{\prime}+1$;\
  \If{$\tau^{\prime}\geq \tau$}
  \State Save the candidate combination(s) with {\color{white}-234}the highest {\color{white}-}number of satisfied{\color{white}-} VT {\color{white}---23}conditions;\
  \State $\tau^{\prime}=0$;\
  \EndIf
  \State $K=Rep$;\
  \If{$K=0$}
  \State $\tau \leftarrow \tau+1$;\
  \State Go to line 2;\
  \Else
  \State Go to line 6;\
  \EndIf
  \State $l=\ell$
  \While{$l> 1$}
  \For{$i\in [1,m^{(\ell-l)}]$}
  \State \hspace{-1mm}Omit parity bits from $\mathbf{\hat{\textbf{x}}}^{(l)}_i$ to get $\mathbf{\hat{\textbf{d}}}^{(l)}_i$;\ 
  \State \hspace{-1mm} $\mathbf{\hat{\textbf{d}}}^{(l)}_{i}=[\mathbf{\hat{\textbf{x}}}^{(l-1)}_i,\mathbf{\hat{\textbf{x}}}^{(l-1)}_{i+1}$,$\ldots,\mathbf{\hat{\textbf{x}}}^{(l-1)}_{i+m-1}]$;\
  \EndFor
  \State $l\leftarrow l-1$;\
  \EndWhile
  \State Decode $\hat{\textbf{d}}^{(1)}_i$ from $\hat{\textbf{x}}^{(1)}_i$ for all $1 \leq i \leq m^{(\ell-1)}$;\
  \State Combine $m^{\ell-1}$ data sections, as $\mathbf{\hat{\textbf{d}}}$.
\end{algorithmic}
\end{multicols}
\end{algorithm}

Algorithm~\ref{algo:decoder} consists of three main parts: In lines 6 to 18, it searches among fragments to attain helpful fragments and candidate combinations. Then, in lines 19 to 28, the algorithm stores candidate combinations with the highest number of satisfied VT conditions and checks the limited memory of the searching algorithm. Finally, it deletes the parity bits from the estimated VT code to decode the data bits. Below, we explain the details.


We initialize the algorithm using $\tau=1$, $K=1$, $\boldsymbol\Gamma_1=\phi$, $\beta_1=S(\textbf{x})$, and $\tau^{\prime}=0$ where $\tau$ represents the limited memory, $K$ is the number of candidate combinations, $\boldsymbol\Gamma_k, 1 \leq k \leq K$ illustrates the $k^{\mathrm{th}}$ candidate combination, $\beta_k, 1 \leq k \leq K$ shows the fragments $\in \{S(\textbf{x})\setminus \boldsymbol\Gamma_k\}$, and $\tau^{\prime}$ is an index that follows the limited memory. Next, we define $Rep$, initialized to zero, to denote the number of new candidate combinations constructed by merging $\boldsymbol\Gamma_k$ and helpful fragments. Then, for each $\boldsymbol\Gamma_k$, we search between all $\mathbf{y}_j \in \beta_k$ to find the helpful fragments and set $\boldsymbol\Gamma^{\prime}=[\boldsymbol\Gamma_k~~\mathbf{y}_j ]$. If $\boldsymbol\Gamma^{\prime}$ is a candidate combination we increase $Rep$ by one, add $\boldsymbol\Gamma^{\prime}$ as a new candidate combination, and consider $\beta_{K+Rep}=$ fragments $\in \{S(\textbf{x})\setminus \boldsymbol\Gamma_{K+Rep}\}$. Then, we check the length of $\boldsymbol\Gamma_{K+Rep}$. If its length is equal to $n$ we say $\hat{\textbf{x}}=\boldsymbol\Gamma_{K+Rep}$ and go to line 29. After checking all candidate combinations, we remove the first $K$ candidate combinations because these $K$ candidate combinations create new candidate combinations with more elements in lines 6 to 16. Then, in line 19, we increase index $\tau^{\prime}$ by one and check whether we have used the whole memory or not. If so, we keep only the candidate combination(s), which provide the highest number of satisfied VT conditions and put $\tau^{\prime}=0$. Next, we use $K=Rep$ as the number of new candidate combinations. In line 24, if $K=0$, we realize that no candidate combination met the VT conditions using limited memory $\tau$. Hence, we set $\tau=\tau+1$ and restart the decoding procedure. Otherwise, we go to line 6 and continue with the remained candidate combinations. 

It is worth mentioning that the searching procedure between fragments is done when the decoder obtains a $\boldsymbol\Gamma_{K+Rep}$ with length $n$. Then, we call $\boldsymbol\Gamma_{K+Rep}$ as $\hat{\textbf{x}}$ and removes its parity bits during lines 29 to 36. Specifically, we set $l=\ell$ and while $l>1$ do the following: For $i^{\mathrm{th}}$ codeword in $l^{\mathrm{th}}$ layer, where $1\leq i \leq m^{(\ell-l)}$, we remove the parity bits from the end of $\hat{\textbf{x}}^{(l)}_i$ to decode $\hat{\textbf{d}}^{(l)}_i$. In the next line, we break $\hat{\textbf{d}}^{(l)}_i$ into $m$ sections and consider each section as a codeword of layer $l-1$. Next, we decrease $l$ by one and repeat lines 31 to 33. Then, we recover $m^{(\ell-1)}$ data sections of layer 1 by just removing the parity bits from corresponding codewords in layer 1. Finally, we merge these $m^{(\ell-1)}$ data sections and obtain $\hat{\textbf{d}}$.

\vspace{-4mm}
\subsection{An example of a nested VT decoder}
\begin{figure}
    \centering
    \includegraphics[trim = 0mm 0mm 0mm 0mm, clip, scale=5, width=0.72\linewidth, draft=false]{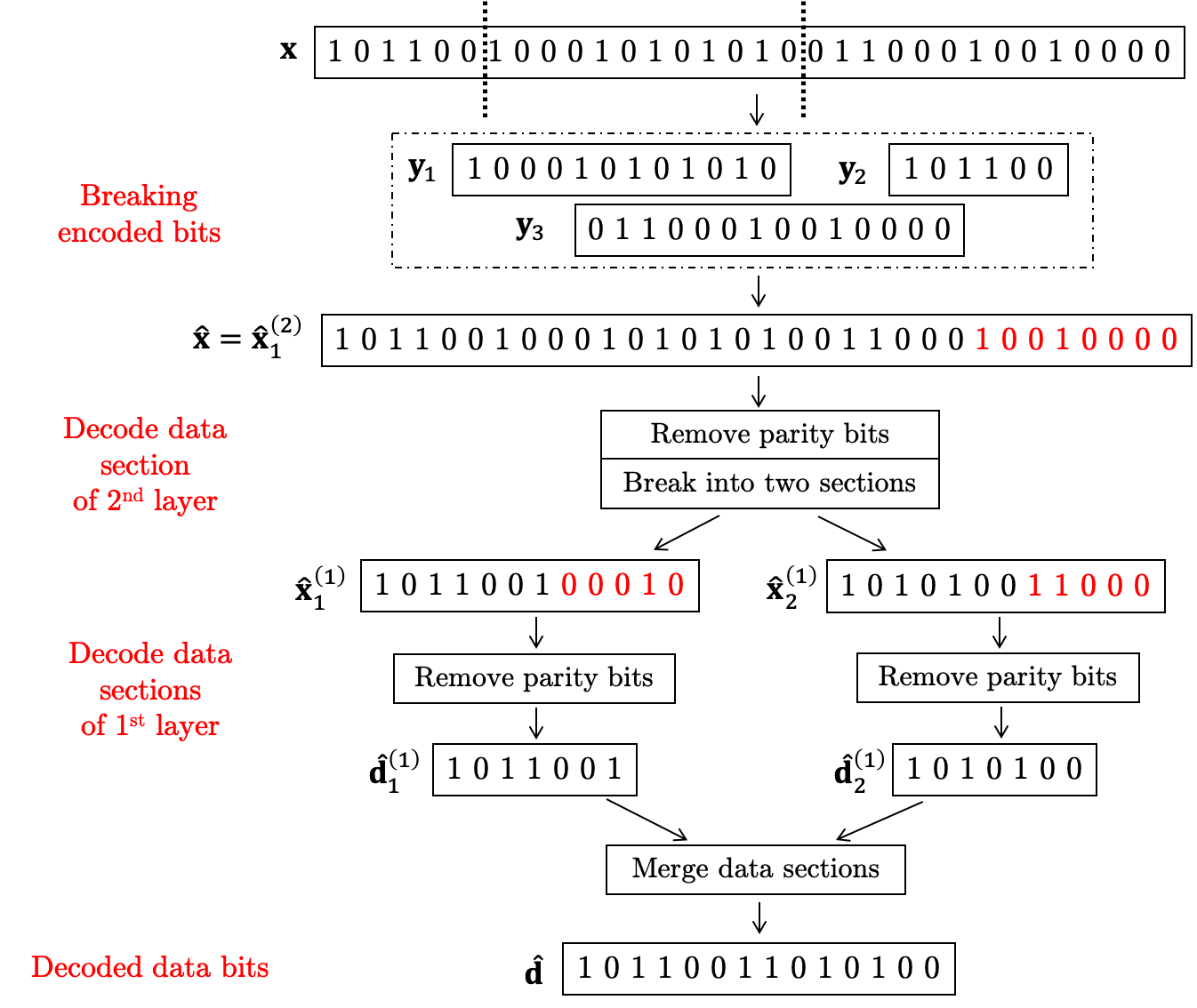}
  \caption{An example of nested VT decoder when $n=32$, $M=3$, $\ell=2$, $m=2$, and $d_{sec}=7$.\label{Fig:ex_nested_decoder}}
  \vspace{-6mm}
\end{figure}

We use the same codeword of Fig.~\ref{Fig:ex_nested_encoder} as nested VT codeword $\textbf{x}$. We consider an example where $\textbf{x}$ goes through the chop-and-shuffle channel and breaks into $M=3$ different pieces as Fig.~\ref{Fig:ex_nested_decoder}. Then, the decoder searches through all permutations to find a string that satisfies all VT conditions. As we can see in Fig.~\ref{Fig:ex_nested_decoder}, the decoder finds $\hat{\textbf{x}}=[1~0~1~1~0~0~1~0~0~0~1~0~1~0~1~0~1~0~0~1~1~0~0~0~1~0~0~1~0~0~0~0]$ as the estimated version of the nested VT codeword. We know that $\hat{\textbf{x}}=\hat{\textbf{x}}_1^{(2)}$. So, the decoder removes parity bits and obtains $\hat{\textbf{d}}_1^{(2)}=[1~0~1~1~0~0~1~0~0~0~1~0~1~0~1~0~1~0~0~1~1~0~0~0]$. Furthermore, we know that $\hat{\textbf{d}}_1^{(2)}$ includes two VT codewords of the first layer. Therefore, we divide $\hat{\textbf{d}}_1^{(2)}$ into two codewords $\hat{\textbf{x}}_1^{(1)}=[1~0~1~1~0~0~1~0~0~0~1~0]$ and $\hat{\textbf{x}}_2^{(1)}=[1~0~1~0~1~0~0~1~1~0~0~0]$. In this stage, we remove parity bits of $\hat{\textbf{x}}_1^{(1)}$ and $\hat{\textbf{x}}_2^{(1)}$ to attain $\hat{\textbf{d}}_1^{(1)}=[1~0~1~1~0~0~1]$ and $\hat{\textbf{d}}_2^{(1)}=[1~0~1~0~1~0~0]$. We know that estimated data bits $\hat{\textbf{d}}$ is created by combining $\hat{\textbf{d}}_1^{(1)}$ and $\hat{\textbf{d}}_2^{(1)}$. As a result, we have $\hat{\textbf{d}}=[1~0~1~1~0~0~1~1~0~1~0~1~0~0]$, which is the correct reconstruction.

\vspace{-4mm}
\subsection{Nested VT decoder plus erasure decoder}
\label{subsect:VTdecoder_erasure_decoder}

As we discussed in Section~\ref{Section:erasure_nested}, when our nested VT decoder discovers multiple candidate combinations with length $n$ that satisfy all the VT conditions, it takes the overlap of these output sequences as the reconstructed sequence. This may result in some missing pieces, which can be viewed as erasures. We add an erasure code during encoding and  an erasure decoder to to handle these missing pieces.



\begin{figure*}[h]
    \centering
    \includegraphics[trim = 0mm 0mm 0mm 0mm, clip, scale=6, width=0.99\linewidth, draft=false]{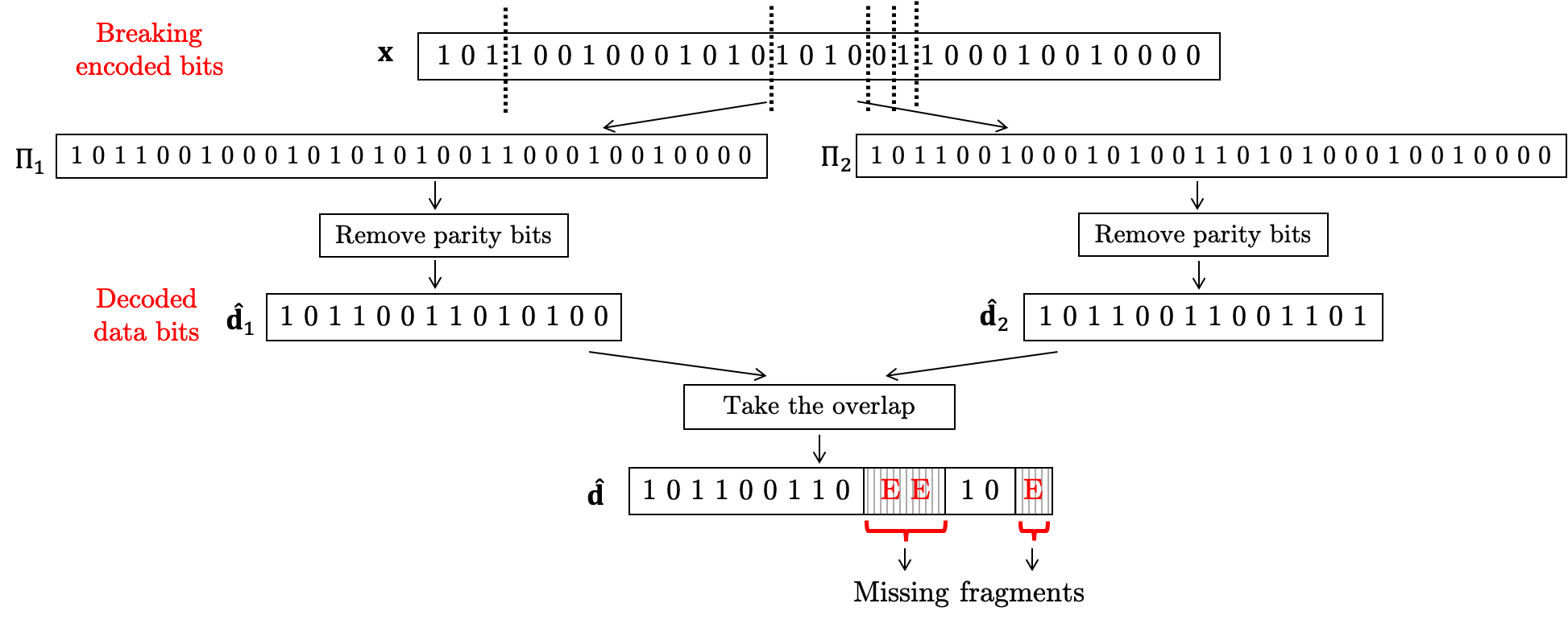}
  \caption{An example of having more than one decoded data bits when $n=32$, $M=6$, $\ell=2$, $m=2$, and $d_{sec}=7$.\label{Fig:ex_overlap}}
  \vspace{-8mm}
\end{figure*}

For instance, in Fig.~\ref{Fig:ex_overlap}, the chop-and-shuffle channel breaks codeword $\textbf{x}=[1~0~1~1~0~0~1~0~0~0~1$ $~0~1~0~1~0~1~0~0~1~1~0~0~0~1~0~0~1~0~0~0~0]$ into $M=6$ different fragments, in which two fragments contain only one bit each. Here, if we use Algorithm~\ref{algo:decoder} to execute the nested VT decoder, we obtain two strings $\Pi_1=[1~0~1~1~0~0~1~0~0~0~1~0~1~0~1~0~1~0~0~1$ $1~0~0~0~1~0~0~1~0~0~0~0]$ and $\Pi_2=[1~0~1~1~0~0~1~0~0~0~1~0~1~0~0~1~1~0~1~0~1~0~0~0~1~0~0~1$ $0~0~0~0]$, both of which meet all the VT conditions and removing the parity bits results in $\hat{\textbf{d}}_1=[1~0~1~1~0~0~1~1~0~1~0~1~0~0]$ and $\hat{\textbf{d}}_2=[1~0~1~1~0~0~1~1~0~0~1~1~0~1]$ as the decoded data bits. This shows that short fragments lead to more than one decoded data sequences. Finally, we take the overlap between $\hat{\textbf{d}}_1$ and $\hat{\textbf{d}}_2$ and conclude that $\hat{\textbf{d}}=[1~0~1~1~0~0~1~1~0~E~E~1~0~E]$ is the recovered data bits where $E$ describes the erasure.

\section{Simulation results}
\label{Section:simulations}
This section provides the simulation results of our work that are done in MATLAB and available online at~\cite{key}. 

We assume input data bits $\textbf{d}$ is a random binary string wherein each element is an independently and identically distributed Bernoulli$(1/2)$ random variable. Moreover, we consider the length of $\textbf{d}$ to be $m^{(\ell-1)}d_{sec}$, and our nested encoder uses $\ell$ encoding layers. Then, we assume the chop-and-shuffle channel breaks codeword $\textbf{x}$ into $M$ different fragments where the length of each broken fragment is an independent random variable following a Geometric distribution with $p_n=\alpha/\log_2(n)$. Furthermore, we use an erasure code with a rate of $1-\epsilon$ before our nested VT encoder to indicate the role of missing pieces in our simulation, and $\epsilon$ is obtained heuristically. Moreover, the rate of our scheme equals to 
\begin{align}
    R_t=(1-\epsilon)R,
\end{align}
where $R$ is the rate of nested VT code. 

To contain complexity and simulation runtime, we limit the decoding time to $\Delta$ and an error occurs if the decoder attains at least one $\hat{\textbf{d}}$ during the given time, but $\hat{\textbf{d}}\neq\textbf{d}$. In Table~\ref{table:increase_Delta}, we compare the number of cases where the decoding algorithm cannot recover any $\hat{\textbf{d}}$ within $\Delta$, and observe that these cases represent a small fraction of total cases and decrease as we increase $\Delta$. The table shows the number of cases where our decoder fails to find any $\hat{\mathbf{d}}$ within $\Delta$ normalized by the total number of simulated cases.

\begin{table}[h]
\caption{Number of cases where our decoder fails to find any $\hat{\mathbf{d}}$ within $\Delta$ for: $d_{sec}=185$, $m=3$, $\ell=3$, $n = 2016$, $\alpha=0.05$, $R_t=0.8194$, and $r_i^{(l)}=0$.}
\centering
\begin{tabular}{|c|c|}
\hline

 $\Delta~(\mathrm{sec})$ & Normalized $\#$ of failed cases\\[0.5ex]
\hline \hline
10 & 0.014\\
50 & 0.007 \\
100 & 0.005\\
500 & 0.002\\ 
\hline
\end{tabular}
\label{table:increase_Delta}
\vspace{-5mm}
\end{table}

\begin{figure}
    \centering
    \includegraphics[trim = 0mm 0mm 0mm 0mm, clip, scale=7, width=0.99\linewidth, draft=false]{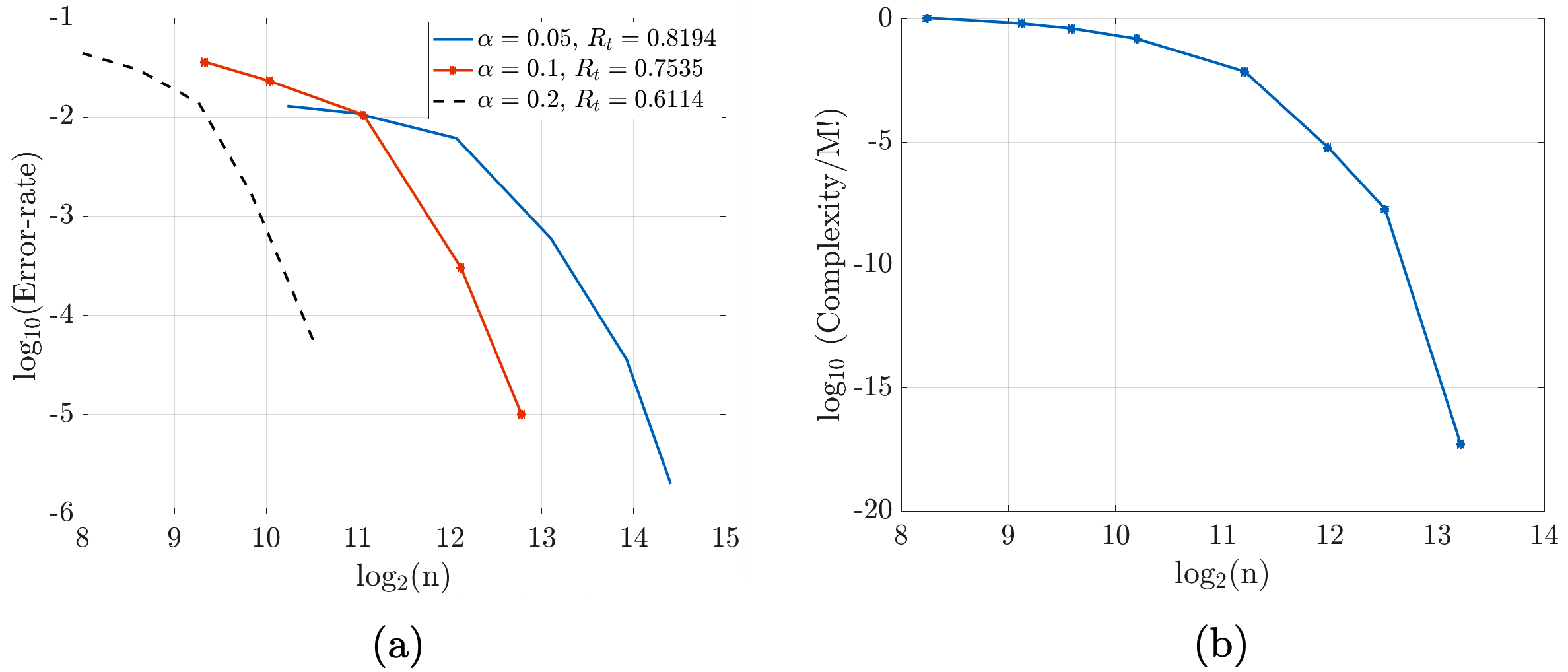}
  \caption{(a) Error rates with different values of $\alpha$; (b) The complexity of our approach in terms of a number of cases that are searched through the decoding algorithm when $\alpha=0.05$, $\ell=3$, and $n \in (2^8,2^{14})$.\label{Fig:error_rates}
  }
  \vspace{-8mm}
\end{figure}

Fig.~\ref{Fig:error_rates}(a) depicts the error rate of our solution as a function of the codeword length $n$ on a logarithmic scale. Specifically, it includes three error rate curves when  $\alpha=0.05, 0.1$, and $0.2$ and $R_t$ is equal to $0.8194, 0.7535,$ and $0.6114$, respectively. As we see in Fig.~\ref{Fig:error_rates}(a), the error rate decreases as the length of codeword increases, and this shows the decoding error decreases as $n$ grows. 

Note that we obtain $\epsilon$
heuristically such that $R_t=(1-\epsilon)R$ holds for all cases.

Earlier we asked whether using either $r_i^{(l)}\neq 0$ or unique residues as $r_i^{(l)}=i-1$ for different VT codewords in each layer would benefit the results. To answer this, in Table~\ref{table:diff_r_methods}, we summarize the simulation results for different methods of selecting $r_i^{(l)}$ when $d_{sec}=185$, $m=3$, $\ell=3$, $n = 2016$, $\alpha=0.05$, $R_t=0.8194$, and $\Delta=20~\mathrm{sec}$. 
According to Table~\ref{table:diff_r_methods}, using different residues for different VT codewords reduces the number of cases where the decoder does not obtain $\hat{\textbf{d}}$ during $\Delta$, but only slightly. However, it shows that using $r_i^{(l)} \neq 0$ degrades the error rate. In fact, \cite{sloane2002single} shows that there are more VT codewords with $r_i^{(l)} = 0$ than any other value for $r_i^{(l)}$. This means with higher values of $r_i^{(l)}$, the number of codewords decreases, and it might be the case that this increases the chance of different arrangements satisfying the VT conditions. 



\begin{table}[h]
\caption{Comparison between different methods of choosing $r_i^{(l)}$ for: $d_{sec}=185$, $m=3$, $\ell=3$, $n = 2016$, $\alpha=0.05$, $R_t=0.8194$, and $\Delta=20~\mathrm{sec}.$}
\centering
\begin{tabular}{|c|c|c|}
\hline

$r_i^{(l)}$ & Normalized $\#$ of failed cases & Error rate\\[0.5ex]

\hline \hline
Unique & 0.006 & 0.011 \\
0  & 0.008 & 0.011 \\
5  & 0.196 & 0.018\\
10 & 0.263 & 0.265\\
50 & 0.267 & 0.286\\ 
\hline
\end{tabular}
\label{table:diff_r_methods}
\vspace{-3mm}
\end{table}

In Fig.~\ref{Fig:error_rates}(b), we describe the complexity of our decoding scheme in terms of the number of permutations that our approach explores during the decoding process, and we use the brute-force search as the baseline. The brute-force solution goes through $M!$ cases, which grows approximately (Stirling's approximation) as $M^M$. 
In Fig.~\ref{Fig:error_rates}(b),  we consider the complexity of our approach divided by $M!$ versus the codeword length in a logarithmic fashion. It indicates that our approach significantly reduces the complexity compared to the brute-force method.
\vspace{-2mm}
\section{Conclusion}
\label{Section:conclusion}
Motivated by recent advances in DNA-based storage, in this paper, we investigated the chop-and-shuffle channel that breaks the input sequence into a random number of variable-length fragments. We designed an encoding/decoding scheme based on a nested structure of Varshamov-Tenengolts (VT) codes to retrieve data bits from the non-overlapping fragments. We showed through simulations that our approach provides higher rates than the lower-bound in~\cite{TornPaperCoding}, and the decoding error decreases as the codeword length increases. Finally, we proposed a new method with linear complexity to build a single VT code that requires fewer parity bits compared to~\cite{sloane2002single}.

\begin{appendices}
\section{Proof of theorem~\ref{theorem_length_parity}}
\label{appndx:thm_length_parity}


\begin{reptheorem}{theorem_length_parity}
Given any data-bit sequence of length $n_d$, it is possible to construct a VT code as defined in (\ref{eq_VT_def}) of length $n$ with residue $0 \leq r \leq n$ where $n = n_d + p$ for $p=\Big\lceil\frac{1+\sqrt{1+8n_d}}{2}\Big\rceil$.
\end{reptheorem}

\begin{proof}
According to \eqref{eq:construct_x}, $\textbf{x}=[\textbf{d} \quad \textbf{p}]$ is the VT codeword where \textbf{d} and \textbf{p} represent the data bits with length $n_d$ and parity bits with length $p$, respectively. We will show later in Appendix~\ref{appndix:find_parity} that we set
\begin{align}
    \label{eq:residue_Delta_appndix}
    \sum_{i^{\prime}=1}^{p}i^{\prime}x_{n+1-i^{\prime}}\equiv \delta, \mod (n+1),
\end{align}
as the residue of parity bits where $0\leq\delta\leq n$. Moreover, we know that $\frac{p(p+1)}{2}$ is the maximum value of $\sum_{i^{\prime}=1}^{p}i^{\prime}x_{n+1-i^{\prime}}$, which happens if all parity bits are equal to 1. As a result, to generate $\delta$, we need
\begin{align}
\label{eq:min_p_quad}
& \frac{p(p+1)}{2}\geq n=n_d+p\\ \nonumber 
& p^2-p\geq2n_d
\end{align}

To solve the quadratic inequality in (\ref{eq:min_p_quad}), we need to find $p$ in the equality case. To do this, we derive the roots of the following quadratic equality
\begin{align}
\label{eq:equality_p}
& p^2-p-2n_d=0 \\ \nonumber
& p=\frac{1\pm \sqrt{1+8n_d}}{2} \rightarrow \left\{
                \begin{array}{ll}
                  p_1= \frac{1- \sqrt{1+8n_d}}{2}\\
                  p_2=\frac{1+ \sqrt{1+8n_d}}{2}.
                \end{array}
            \right.
\end{align}
We can show that (\ref{eq:min_p_quad}) is feasible when $p\leq p_1$ or $p\geq p_2$. Additionally, since $p$ is a positive integer number, $p\leq p_1$ is not acceptable. As a result, we use $p=\ceil{p_2}=\ceil{\frac{1+\sqrt{1+8n_d}}{2}}$ as the minimum integer number that satisfy $p\geq p_2$, and this completes the proof. 


\end{proof}
\vspace{-8mm}
\section{Finding parity bits $\textbf{p}$}
\label{appndix:find_parity}
In this section, we demonstrate how to obtain parity bits $\textbf{p}$ with length $p=\ceil{\frac{1+\sqrt{1+8n_d}}{2}}$ to ensure that codeword $\textbf{x}$ meets desired residue $0 \leq r \leq n$. To do this, we define $r^{\prime}$ as
\begin{align}
    \sum_{i=1}^{n_d}id_i\equiv r^{\prime} \mod (n+1),
\end{align}
where $d_i$ is the $\text{i}^{\mathrm{th}}$ element of $\textbf{d}$. Moreover, we know that $\textbf{d}$ represents the first $n_d$ elements of $\textbf{x}$. Therefore, we have,
\begin{align}
\label{eq:sum_residue}
\sum_{i=1}^{n}ix_i&=\sum_{i=1}^{n_d}ix_i+\sum_{i=n_d+1}^{n}ix_i=\beta(n+1)+r^{\prime}+\sum_{i=n_d+1}^{n}ix_i\\ \nonumber 
&\hspace{-17pt}\overset{(i^{\prime} \triangleq n+1-i)}{=}\beta(n+1)+r^{\prime}+\sum_{i^{\prime}=1}^{n-n_d}(n+1-i^{\prime})x_{n+1-i^{\prime}}\\ \nonumber
&= \underbrace{\beta(n+1)}_{a_1}+r^{\prime}+\underbrace{(n+1)\sum_{i^{\prime}=1}^{p}x_{n+1-i^{\prime}}}_{a_2}-\sum_{i^{\prime}=1}^{p}i^{\prime}x_{n+1-i^{\prime}}.
\end{align}
Then, we use \eqref{eq:sum_residue} to calculate the residue. We know that the residues of $a_1$ and $a_2$ in \eqref{eq:sum_residue} equal to zero. As a result, we have,
\begin{align}
    \label{eq:residue_i_xi}
    \sum_{i=1}^{n}ix_i\equiv r^{\prime}-\delta, \mod (n+1),
\end{align}
where $\delta$ is equal to 
\begin{align}
    \label{eq:residue_Delta}
    \sum_{i^{\prime}=1}^{p}i^{\prime}x_{n+1-i^{\prime}}\equiv \delta, \mod (n+1).
\end{align}
Here, \eqref{eq:residue_Delta} shows that by calculating the residue of codeword $\textbf{x}$ from the right-hand side (starting from the parity bits), $\delta$ represents the residue of parity bits string from the right-hand side. Therefore, to achieve desired residue $r$, all we need is to create parity bits $\textbf{p}$ such that $\delta$ equals to
\begin{equation}
\delta=\left\{
                \begin{array}{ll}
                  r^{\prime}-r \quad \quad \quad \quad \quad \quad ~r^{\prime}\geq r,\\
                  (n+1)-(r-r^{\prime}) \quad r^{\prime}< r.
                \end{array}
            \right.
\end{equation}

To derive parity bits $\textbf{p}$, we propose a four-step procedure as below: 

{\it Step 1 $\rightarrow$ Define $k=p-i^{\prime}+1$}: We use $k$ to show the position of parity bit which starts from left side of parity bits;

{\it Step 2 $\rightarrow$ Solve $pk-\frac{k(k-1)}{2}\leq \delta$:} We solve this inequality and obtain the largest non-negative integer value of $k$ as $k_{max}$. Here, it is essential to show that $pk-\frac{k(k-1)}{2}\leq \delta$ is feasible for $p\geq 1$. To do this, we have
\begin{align}
\label{eq:j_max_inequal}
& pk-\frac{k(k-1)}{2}\leq \delta \overset{(a)}{\leq} \frac{p^2+p}{2} \\ \nonumber
& k^2-(2p+1)k+p^2+p\geq 0,
\end{align}
where $(a)$ holds true since the maximum value of $\delta$ is equal to $(p^2+p)/2$ when all bits of $\textbf{p}$ are equal to 1.
To solve the quadratic inequality in (\ref{eq:j_max_inequal}), we need to find $k$ in the equality case. More precisely, we derive the roots of the following quadratic equality
\begin{align}
\label{eq:j_max_equal}
& k^2-(2p+1)k+p^2+p=0 \\ \nonumber
& k=\frac{2p+1\pm 1}{2} \rightarrow \left\{
                \begin{array}{ll}
                  k_1= \frac{2p}{2}=p\\
                  k_2=\frac{2p+2}{2}=p+1.
                \end{array}
            \right.
\end{align}



Thus, based on $k_1$ and $k_2$ in \eqref{eq:j_max_equal}, we know that either $k_{max}\leq k_1=p$ or $k_{max}\geq k_2=p+1$ satisfies the inequality in (\ref{eq:j_max_inequal}). Moreover, we know $0\leq k_{max}\leq p$ and $k_{max}$ at most can be equal to $p$. As the result, it shows that always there is $0\leq k_{max}\leq p$, which satisfies the quadratic inequality in (\ref{eq:j_max_inequal}), and this proves that $pk-\frac{k(k-1)}{2}\leq \delta$ is feasible for $p\geq 1$.

{\it Step 3 $\rightarrow$ Put all parity bits in positions $1\leq k \leq k_{max}$ equal to 1 when $k_{max}\geq 1$:} In this step, we create an initial version of parity bits by placing one in position(s) $1\leq k \leq k_{max}$. Consequently, we have an initial parity bits such that $\sum_{i^{\prime}=1}^{p}i^{\prime}x_{n+1-i^{\prime}}=pk_{max}-\frac{k_{max}(k_{max}-1)}{2}$;

{\it Step 4 $\rightarrow$ Place parity bit in position $b$ where $b \triangleq\delta-pk_{max}-\frac{k_{max}(k_{max}-1)}{2}$:} From {\it Step 3}, we know that
\begin{align}
\label{eq:p_k_max_less_delta}
    \sum_{i^{\prime}=1}^{p}i^{\prime}x_{n+1-i^{\prime}}=pk_{max}-\frac{k_{max}(k_{max}-1)}{2}\leq \delta.
\end{align} 
Therefore, to convert the inequality to equality in \eqref{eq:p_k_max_less_delta}, we add $b=\delta-pk_{max}-\frac{k_{max}(k_{max}-1)}{2}$ to $\sum_{i^{\prime}=1}^{p}i^{\prime}x_{n+1-i^{\prime}}$ in \eqref{eq:p_k_max_less_delta}. To do this, we put 1 in the position of $b$.

\vspace{-4mm}
\section{Complexity of finding parity bits $\textbf{p}$}
\label{appndix:complexity_parity}

In this appendix, we describe the complexity of our strategy to find parity bits. To explain the details, we divide the complexity into the following three parts:

{\it (i) $\sum_{i=1}^{n} ix_i$:} Here, we need to perform $(n-1)$ additions to get $\sum_{i=1}^{n} ix_i$. We note that we do not need multiplications as $x_i \in \{0,1\}$, which simply means add or do not add $i$. Furthermore, each addition is a constant-time operation. Thus, $\mathcal{O}(n)$ is the complexity of $\sum_{i=1}^{n} ix_i$;

{\it (ii) Mod function:} The author in \cite{shoup2009computational}, 
shows that the computing complexity of mod function of $r\equiv \sum_{i=1}^{n} ix_i ~~\text{mod}(n+1)$ is $\mathcal{O}(\text{length}(n+1) \times \text{length}(q))$ where $q$ satisfies \begin{align}
   \sum_{i=1}^{n} ix_i=q(n+1)+r, 
\end{align}
and $\text{length}(z)=\lfloor\log_2(z)\rfloor+1$ describes the length of number $z$ in binary domain where $\lfloor.\rfloor$ is the floor function. 
We know that $q\leq \frac{n}{2}$ because $\sum_{i=1}^{n} ix_i\leq \frac{n(n+1)}{2}$. Hence, the computing complexity of mod function is $\mathcal{O}(\text{length}(n+1)\times\text{length}(\frac{n}{2}))= \mathcal{O}(\Big[\lfloor\log_2(n+1)\rfloor+1 \Big] \times \Big[\lfloor\log_2\frac{n}{2}\rfloor+1 \Big])$;

{\it (iii) Solving quadratic inequality:} In \cite{kim1994polynomial}, the authors explain that the complexity of finding the roots of a quadratic polynomial with $\epsilon$-error is equal to 
\begin{align}
    \mathcal{O}(2(\log_{10}2)^2|\log_{10}\epsilon|+4(\log_{10}2)^2).
\end{align}
We note that since all calculations are done in the binary domain, we need to take into account the complexity of converting the root of the quadratic equation into the binary domain, which is given as
\begin{align}
    \mathcal{O}(\lfloor \log_2 k_{max} \rfloor +1)\overset{(a)}{\leq}\mathcal{O}(\lfloor \log_2 p \rfloor +1)\overset{(b)}{<}\mathcal{O}(\lfloor \log_2 n \rfloor +1),
\end{align}
where $(a)$ and $(b)$ hold true since $k_{max}\leq p<n$. As a result, the overall complexity of finding the root of the quadratic equation equals to $\mathcal{O}(\lfloor \log_2 n \rfloor +1)$.

According to the above results, we conclude that the complexity of our encoding method is also linear (i.e., $\mathcal{O}(n)$), similar to Sloane's approach in \cite{sloane2002single}. 
\vspace{-4mm}
\section{Proof of theorem~\ref{theorem:encoding_rate}}
\label{appndix_proof_encoding_rate}

\begin{reptheorem}
{theorem:encoding_rate}
Given that $n_d=m^{(\ell-1)}d_{sec}$ is the length of data bits $\textbf{d}$ and for $d_{sec}\geq 36$, the encoding rate of the $\ell$-layered nested VT code is bounded by 
\begin{align}
R^-<R<R^+,
\end{align}
where 
\begin{align}
    R^-=\frac{m^{\ell-1}d_{sec}}{\Big(m^{(\frac{\ell-1}{2})}\sqrt{d_{sec}}+\frac{\sqrt{2.5}}{2}\frac{m^{\frac{\ell}{2}}-1}{\sqrt{m}-1}\Big)^2},
\end{align}
and the upper bound, $R^+$, is equal to
\begin{align}
    R^+=\frac{2m^{\ell-1}d_{sec}}{\Big(m^{(\frac{\ell-1}{2})}\sqrt{2d_{sec}}+\frac{m^{\frac{\ell}{2}}-1}{\sqrt{m}-1}\Big)^2}.
\end{align}
\end{reptheorem}

\begin{proof}
Our nested VT code with $\ell$ layers converts the input data bits $\textbf{d}$ with length $m^{(\ell-1)}d_{sec}$ to codeword $\textbf{x}$ with length $n$. Hence, the encoding rate is
\begin{align}
\label{eq_R_general}
    R=\frac{m^{(\ell-1)}d_{sec}}{n},
\end{align}

To obtain the lower and upper bounds on $R$, we need to find the lower and upper bounds of $n$. Therefore, we divide the proof into two parts:

$\bullet~\textbf{Lower bound on n}$

According to Theorem~\ref{theorem_length_parity}, the length of each codeword in the first layer is equal to
\begin{align}
\label{tilda_n_1}
    \Tilde{n}_1=d_{sec}+\Bigg\lceil\frac{1+\sqrt{1+8d_{sec}}}{2}\Bigg\rceil,
\end{align}
Next, we have
\begin{align}
\label{eq:lower_tilda_n_1}
  \Tilde{n}_1&\overset{(a)}{\geq}d_{sec}+\frac{1+\sqrt{1+8d_{sec}}}{2}>d_{sec}+\frac{1+\sqrt{8d_{sec}}}{2}=\frac{(\sqrt{2d_{sec}}+1)^2}{2},
\end{align}
where $(a)$ is correct because $\ceil{u}\geq u$ for any real number $u$. Then, we define $\Tilde{d}_{sec}\triangleq\sqrt{2d_{sec}}$ and rewrite \eqref{eq:lower_tilda_n_1} as
\begin{align}
   \Tilde{n}_1>\frac{(\Tilde{d}_{sec}+1)^2}{2}.
\end{align}
Moreover, $\Tilde{n}_2$ contains $m$ different sections of $\Tilde{n}_1$, which means
\begin{align}
\label{eq:n2_tilde}
   \Tilde{n}_2=m\Tilde{n}_1+\Bigg\lceil\frac{1+\sqrt{1+8m\Tilde{n}_1}}{2}\Bigg\rceil. 
\end{align}
Similar to \eqref{eq:lower_tilda_n_1}, we have 
\begin{align}
    \Tilde{n}_2&>\frac{(\sqrt{2m\Tilde{n}_1}+1)^2}{2}=\frac{(\sqrt{m}(\Tilde{d}_{sec}+1)+1)^2}{2}.
\end{align}
Then, following the same rule, we can show that
\begin{align}
\label{eq_lower_n_L}
   \Tilde{n}_{\ell}&>\frac{(\overbrace{\sqrt{m}(\sqrt{m} \ldots \sqrt{m} }^{\ell-1~ \text{coefficients}}(\Tilde{d}_{sec}+1)\overbrace{+1)+1)\ldots )+1}^{\ell-1~ \text{coefficients}})^2}{2}\\ \nonumber
   &=\frac{(m^{(\frac{\ell-1}{2})}\Tilde{d}_{sec}+m^{(\frac{\ell-1}{2})}+m^{(\frac{\ell-2}{2})}+\ldots+\sqrt{m}+1)^2}{2}\\ \nonumber
   &=\frac{1}{2}(m^{(\frac{\ell-1}{2})}\Tilde{d}_{sec}+\frac{m^{\frac{\ell}{2}}-1}{\sqrt{m}-1})^2=\frac{1}{2}(m^{(\frac{\ell-1}{2})}\sqrt{2d_{sec}}+\frac{m^{\frac{\ell}{2}}-1}{\sqrt{m}-1})^2,
\end{align}
which represents a lower bound on $n$ since the nested VT code in $\ell^{\mathrm{th}}$ layer includes only one section (i.e., $n=\Tilde{n}_{\ell}$).

$\bullet~\textbf{Upper bound on n}$

To derive an upper bound on $n$, we have
\begin{align}
\label{simp_tilda_n_1}
  \Tilde{n}_1&\overset{(b)}{<}(d_{sec}+1)+\frac{1+\sqrt{1+8d_{sec}}}{2}<(d_{sec}+1)+0.5+\sqrt{2(d_{sec}+1)}\\ \nonumber
  &=(\sqrt{d_{sec}+1}+\frac{\sqrt{2}}{2})^2\overset{(c)}{<}(\sqrt{d_{sec}}+\frac{\sqrt{2.5}}{2})^2,
\end{align}
where $(b)$ holds true since $\ceil{u}<u+1$ for any real number $u$. Then, to show for which range of $ d_ {sec}$, $ (c)$ is true, we extend the inequality in \eqref{simp_tilda_n_1} as
\begin{align}
\label{eq:dsec_ineq_1}
    d_{sec}+\frac{3}{2}+\sqrt{2(d_{sec}+1)}&<d_{sec}+0.625+\sqrt{2.5d_{sec}}\\ \nonumber
    0.875+\sqrt{2(d_{sec}+1)}&<\sqrt{2.5d_{sec}},
\end{align}
where both sides of \eqref{eq:dsec_ineq_1} are positive numbers, and doubling them does not change the inequality. Thus, we have
\begin{align}
\label{eq:dsec_ineq_2}
    (0.875+\sqrt{2(d_{sec}+1)})^2&<(\sqrt{2.5d_{sec}})^2\\ \nonumber
    2.47\sqrt{(d_{sec}+1)}&<0.5d_{sec}-2.766 \\ \nonumber
    (2.47\sqrt{(d_{sec}+1)})^2&\overset{(d)}{<}(0.5d_{sec}-2.766)^2 \\ \nonumber
    -d_{sec}^2+35.464d_{sec}-6.2&<0.
\end{align}
where $(d)$ happens because two sides of inequality are positive numbers. Then, by solving \eqref{eq:dsec_ineq_2}, we obtain that $d_{sec}>35.288$, which means that condition $(c)$ in \eqref{simp_tilda_n_1} holds true if $d_{sec}\geq 36$. 

Finally, following the same steps in deriving the lower bound of $n$ leads to
\begin{align}
\label{eq_upper_n_L}
   \Tilde{n}_{\ell}
   <(m^{(\frac{\ell-1}{2})}\sqrt{d_{sec}}+\frac{\sqrt{2.5}}{2}\frac{m^{\frac{\ell}{2}}-1}{\sqrt{m}-1})^2.
\end{align}
Here, we use \eqref{eq_R_general}, \eqref{eq_lower_n_L}, and \eqref{eq_upper_n_L} to conclude that
\begin{align}
R^{-}<R<R^+,
\end{align}
where 
\begin{align}
    R^-=\frac{m^{\ell-1}d_{sec}}{\Big(m^{(\frac{\ell-1}{2})}\sqrt{d_{sec}}+\frac{\sqrt{2.5}}{2}\frac{m^{\frac{\ell}{2}}-1}{\sqrt{m}-1}\Big)^2},
\end{align}
and
\begin{align}
    R^+=\frac{2m^{\ell-1}d_{sec}}{\Big(m^{(\frac{\ell-1}{2})}\sqrt{2d_{sec}}+\frac{m^{\frac{\ell}{2}}-1}{\sqrt{m}-1}\Big)^2},
\end{align}
which completes the proof.
\end{proof}

\end{appendices}

{\footnotesize
\bibliographystyle{ieeetr}
\bibliography{refs.bib}

\begin{thebibliography}{10}

\bibitem{church2012next}
G.~M. Church, Y.~Gao, and S.~Kosuri, ``Next-generation digital information
  storage in {DNA},'' {\em Science}, vol.~337, no.~6102, pp.~1628--1628, 2012.

\bibitem{goldman2013towards}
N.~Goldman, P.~Bertone, S.~Chen, C.~Dessimoz, E.~M. LeProust, B.~Sipos, and
  E.~Birney, ``Towards practical, high-capacity, low-maintenance information
  storage in synthesized {DNA},'' {\em Nature}, vol.~494, no.~7435, pp.~77--80,
  2013.

\bibitem{grass2015robust}
R.~N. Grass, R.~Heckel, M.~Puddu, D.~Paunescu, and W.~J. Stark, ``Robust
  chemical preservation of digital information on {DNA} in silica with
  error-correcting codes,'' {\em Angewandte Chemie International Edition},
  vol.~54, no.~8, pp.~2552--2555, 2015.

\bibitem{yazdi2015rewritable}
S.~H.~T. Yazdi, Y.~Yuan, J.~Ma, H.~Zhao, and O.~Milenkovic, ``A rewritable,
  random-access {DNA}-based storage system,'' {\em Scientific reports}, vol.~5,
  no.~1, pp.~1--10, 2015.

\bibitem{erlich2017dna}
Y.~Erlich and D.~Zielinski, ``{DNA} fountain enables a robust and efficient
  storage architecture,'' {\em Science}, vol.~355, no.~6328, pp.~950--954,
  2017.

\bibitem{organick2018random}
L.~Organick, S.~D. Ang, Y.-J. Chen, R.~Lopez, S.~Yekhanin, K.~Makarychev, M.~Z.
  Racz, G.~Kamath, P.~Gopalan, B.~Nguyen, {\em et~al.}, ``Random access in
  large-scale {DNA} data storage,'' {\em Nature biotechnology}, vol.~36, no.~3,
  pp.~242--248, 2018.

\bibitem{motahari2013information}
A.~S. Motahari, G.~Bresler, and N.~David, ``Information theory of {DNA} shotgun
  sequencing,'' {\em IEEE Transactions on Information Theory}, vol.~59, no.~10,
  pp.~6273--6289, 2013.

\bibitem{bornholt2016dna}
J.~Bornholt, R.~Lopez, D.~M. Carmean, L.~Ceze, G.~Seelig, and K.~Strauss, ``A
  {DNA}-based archival storage system,'' in {\em Proceedings of the
  Twenty-First International Conference on Architectural Support for
  Programming Languages and Operating Systems}, pp.~637--649, 2016.

\bibitem{branton2010potential}
D.~Branton, D.~W. Deamer, A.~Marziali, H.~Bayley, S.~A. Benner, T.~Butler,
  M.~Di~Ventra, S.~Garaj, A.~Hibbs, X.~Huang, {\em et~al.}, ``The potential and
  challenges of nanopore sequencing,'' {\em Nanoscience and technology: A
  collection of reviews from Nature Journals}, pp.~261--268, 2010.

\bibitem{bleidorn2016third}
C.~Bleidorn, ``Third generation sequencing: technology and its potential impact
  on evolutionary biodiversity research,'' {\em Systematics and biodiversity},
  vol.~14, no.~1, pp.~1--8, 2016.

\bibitem{jain2016oxford}
M.~Jain, H.~E. Olsen, B.~Paten, and M.~Akeson, ``The oxford nanopore minion:
  delivery of nanopore sequencing to the genomics community,'' {\em Genome
  biology}, vol.~17, no.~1, pp.~1--11, 2016.

\bibitem{shomorony2020communicating}
I.~Shomorony and A.~Vahid, ``Communicating over the torn-paper channel,'' in
  {\em GLOBECOM 2020-2020 IEEE Global Communications Conference}, pp.~1--6,
  IEEE, 2020.

\bibitem{TornPaperCoding}
I.~Shomorony and A.~Vahid, ``Torn-paper coding,'' {\em IEEE Transactions on
  Information Theory}, 2021.

\bibitem{kiah2016codes}
H.~M. Kiah, G.~J. Puleo, and O.~Milenkovic, ``Codes for {DNA} sequence
  profiles,'' {\em IEEE Transactions on Information Theory}, vol.~62, no.~6,
  pp.~3125--3146, 2016.

\bibitem{ravi2021coded}
A.~N. Ravi, A.~Vahid, and I.~Shomorony, ``Coded shotgun sequencing,'' {\em
  arXiv preprint arXiv:2110.02868}, 2021.

\bibitem{shomorony2019capacity}
I.~Shomorony and R.~Heckel, ``Capacity results for the noisy shuffling
  channel,'' in {\em 2019 IEEE International Symposium on Information Theory
  (ISIT)}, pp.~762--766, IEEE, 2019.

\bibitem{shomorony2021dna}
I.~Shomorony and R.~Heckel, ``{DNA}-based storage: Models and fundamental
  limits,'' {\em IEEE Transactions on Information Theory}, vol.~67, no.~6,
  pp.~3675--3689, 2021.

\bibitem{lenz2019coding}
A.~Lenz, P.~H. Siegel, A.~Wachter-Zeh, and E.~Yaakobi, ``Coding over sets for
  {DNA} storage,'' {\em IEEE Transactions on Information Theory}, vol.~66,
  no.~4, pp.~2331--2351, 2019.

\bibitem{bresler2013optimal}
G.~Bresler, M.~Bresler, and D.~Tse, ``Optimal assembly for high throughput
  shotgun sequencing,'' in {\em BMC bioinformatics}, vol.~14, pp.~1--13, BioMed
  Central, 2013.

\bibitem{nassirpour2020embedded}
S.~Nassirpour and A.~Vahid, ``Embedded codes for reassembling non-overlapping
  random {DNA} fragments,'' {\em IEEE Transactions on Molecular, Biological and
  Multi-Scale Communications}, vol.~7, no.~1, pp.~40--50, 2020.

\bibitem{varshamov1965codes}
R.~Varshamov and G.~Tenengolts, ``Codes which correct single asymmetric errors
  (in russian),'' {\em Automatika i Telemkhanika}, vol.~161, no.~3,
  pp.~288--292, 1965.

\bibitem{sloane2002single}
N.~J. Sloane, ``On single-deletion-correcting codes. codes and designs,'' in
  {\em Proceedings of a Conference Honoring Professor Dijen K. Ray-Chaudhuri on
  the Occasion of His 65th Birthday, Ohio State University}, 2002.

\bibitem{levenshtein1966binary}
V.~I. Levenshtein {\em et~al.}, ``Binary codes capable of correcting deletions,
  insertions, and reversals,'' in {\em Soviet physics doklady}, vol.~10,
  pp.~707--710, Soviet Union, 1966.

\bibitem{helberg2002multiple}
A.~S. Helberg and H.~C. Ferreira, ``On multiple insertion/deletion correcting
  codes,'' {\em IEEE Transactions on Information Theory}, vol.~48, no.~1,
  pp.~305--308, 2002.

\bibitem{abdel2011helberg}
K.~A. Abdel-Ghaffar, F.~Paluncic, H.~C. Ferreira, and W.~A. Clarke, ``On
  helberg's generalization of the levenshtein code for multiple
  deletion/insertion error correction,'' {\em IEEE Transactions on Information
  Theory}, vol.~58, no.~3, pp.~1804--1808, 2011.

\bibitem{abroshan2019coding}
M.~Abroshan, R.~Venkataramanan, L.~Dolecek, and A.~G. i~Fabregas, ``Coding for
  deletion channels with multiple traces,'' in {\em 2019 IEEE International
  Symposium on Information Theory (ISIT)}, pp.~1372--1376, IEEE, 2019.

\bibitem{mappouras2019greenflag}
G.~Mappouras, A.~Vahid, R.~Calderbank, and D.~J. Sorin, ``Greenflag: Protecting
  3d-racetrack memory from shift errors,'' in {\em 2019 49th Annual IEEE/IFIP
  International Conference on Dependable Systems and Networks (DSN)},
  pp.~1--12, IEEE, 2019.

\bibitem{vahid2017correcting}
A.~Vahid, G.~Mappouras, D.~J. Sorin, and R.~Calderbank, ``Correcting two
  deletions and insertions in racetrack memory,'' {\em arXiv preprint
  arXiv:1701.06478}, 2017.

\bibitem{key}
S.~Nassirpour, I.~Shomorony, and A.~Vahid, ``Simulation codes of the reassembly
  codes for chop-and-shuffle channel,'' {\em Available online:
  https://github.com/SajjadNassirpour/Reassembly-Codes-for-Chop-and-Shuffle-Channel},
  2021.

\bibitem{shoup2009computational}
V.~Shoup, {\em A computational introduction to number theory and algebra}.
\newblock Cambridge university press, 2009.

\bibitem{kim1994polynomial}
M.-H. Kim and S.~Sutherland, ``Polynomial root-finding algorithms and branched
  covers,'' {\em SIAM Journal on Computing}, vol.~23, no.~2, pp.~415--436,
  1994.

\end{thebibliography}
}

\end{document}